\documentclass[letterpaper,journal]{IEEEtran}

\usepackage{cite}
\usepackage{amsmath,amssymb,amsfonts,amsthm}
\usepackage{graphicx}
\usepackage{textcomp}
\usepackage{array}
\usepackage{multirow}
\usepackage{enumitem}
\usepackage{algorithm}
\usepackage{algpseudocode}
\usepackage{stfloats}
\usepackage{placeins}
\usepackage{balance}
\usepackage{url}
\usepackage{adjustbox}
\usepackage[hidelinks]{hyperref}
\hypersetup{pdftitle={Regret for Age-of-Information Scheduling under Non-Stationary Channels},
pdfauthor={Xiayi Wang, Jianan Zhang, Xiang Cheng, Yuping Zhao},
pdfkeywords={Age of information, wireless scheduling, drifting bandits, online learning, finite-state Markov decision processes}}
\newtheorem{remark}{Remark}
\newtheorem{theorem}{Theorem}
\newtheorem{lemma}{Lemma}
\newtheorem{proposition}{Proposition}
\newtheorem{corollary}{Corollary}
\newtheorem{assumption}{Assumption}
\def\BibTeX{{\rm B\kern-.05em{\sc i\kern-.025em b}\kern-.08em
		T\kern-.1667em\lower.7ex\hbox{E}\kern-.125emX}}

\DeclareMathOperator*{\argmax}{arg\,max}

\newcommand{\modeldiagramwidth}{0.60\linewidth}
\newcommand{\widediagramwidth}{0.70\linewidth}
\newcommand{\simulationfigurewidth}{0.80\linewidth}
\newcommand{\algorithmtextformat}{\normalfont\normalsize}
\newcommand{\proofstep}[2]{%
  \par\smallskip
  \noindent\textbf{Step #1: #2.}\enspace\ignorespaces
}

\begin{document}
\title{Regret for Age-of-Information Scheduling under Non-Stationary Channels\\
}

\author{Xiayi Wang, Jianan Zhang, Xiang Cheng, and Yuping Zhao

\thanks{The authors are with the State Key Laboratory of Photonics and Communications, School of Electronics, Peking University, China. This work was supported in part by the National Natural Science Foundation of China under Grants 62301011 and 62125101; and in part by the New Cornerstone Science Foundation through the Xplorer Prize.}}

\maketitle

\begin{abstract}
We study scheduling over a horizon of \(T\) slots to minimize the
Age-of-Information (AoI) regret across sources at a receiver. The channel
service rates, namely the transmission-success probabilities, are unknown,
vary over time, and have total variation bounded by \(V_T\).
We consider single- and multi-source systems, scheduling at most one source
per slot. We show that AoI regret and the reward regret of channel selection
have the same order up to a logarithmic factor for drifting
channels. This relation gives an
AoI-regret lower bound and a REXP3 upper bound separated only by logarithmic
factors. We also characterize AoI-regret bounds with
explicit dependence on the number of sources $M$. In the decoupled
multi-source system, the scheduler selects a source and a channel separately.
We extend
the single-source system results to this setting and propose Max-Age~+~REXP3,
whose upper bound matches the lower bound up to logarithmic factors for fixed
$M$. In the coupled multi-source system where each
source is tied to its own channel, we formulate the scheduling problem as a Markov decision process.
Against the finite-horizon oracle, we prove that an optimistic
MDP scheduler has a regret bound with the same $T^{2/3}$ and $V_T^{1/3}$
dependence as in the first two models.
For the single-source and decoupled multi-source systems, we also provide regret
bounds for algorithms that do not need $V_T$ in advance.
\end{abstract}


\section{Introduction}

\IEEEPARstart{R}{eal-time} monitoring, sensing, and control systems rely on timely status updates. In applications such as autonomous driving, industrial automation, remote estimation, and multi-agent coordination, outdated status information may lead to poor control or inference decisions. AoI, the time elapsed since the generation of the most recently received update, has been widely used to quantify information freshness in communication and networked control systems~\cite{kaul2012realtime,yates2021age}.

Wireless channel service rates vary over time because of mobility, fading,
shadowing, interference, and traffic load. We write \(\mu_k(t)\) for the
transmission-success probability of channel \(k\) in slot \(t\). The scheduler
does not know these service rates and estimates them from past transmission
successes and failures. Over the \(T\)-slot horizon, we assume that their
cumulative change is at most \(V_T\). Unlike a switching budget, which only
counts abrupt changes, this variation budget also describes gradual changes.

In an ordinary bandit, choosing a worse channel loses reward only in the
current slot. In an AoI system, a missed update keeps the age larger until
a later successful transmission. Thus, a one-slot channel loss can induce
AoI loss over multiple subsequent slots.

We consider three models. In the single-source model,
the scheduler chooses a channel. In the decoupled multi-source model, an action
chooses one of $M$ sources and one of \(K\) channels separately, and the selected channel
has service rate \(\mu_k(t)\) regardless of the source. The channel determines
whether delivery succeeds, while the source choice determines which age can
reset. For example, a roadside unit can send fresh packets from several sensors
over frequency bands whose service rates do not depend on the sensor. In the
coupled multi-source model, there are \(M\) source-specific channels. Action \(i\) selects
source \(i\) together with channel \(i\), so \(K=M\) in this model. For example, 
a receiver may collect updates from several sensors, each through its own 
source-specific wireless link, so selecting a sensor also determines the channel 
used for transmission.

These action structures also determine the comparators. The single-source
oracle uses the channel with the largest service rate in each slot. We later
prove that the oracle in the decoupled multi-source model combines this
channel choice with Max-Age source scheduling. The oracle in the coupled
multi-source model is a finite-horizon MDP policy: it
knows all channel service rates but
chooses each action before observing the current transmission outcome. It
cannot in general select only the channel with the largest service rate because
different actions reset different source ages. For example, with source ages 20 and 1
and their corresponding channel service rates 0.7 and 0.8, serving the older source can yield a larger
expected age reduction. The oracle therefore uses both service rates and ages.

\subsection{Related Work}

\textit{AoI scheduling under known parameters.}
Foundational work on AoI scheduling under known parameters introduced AoI as a freshness metric and
characterized when to sample or transmit under a given queueing or delay
model~\cite{kaul2012realtime,sun2017update,yates2021age}. Our single-source
model retains the same reset-and-increment age dynamics but adds the need to
learn among \(K\) drifting channels. For multiple sources, one line of work
selects which update to serve over a limited transmission opportunity, with
objectives involving correlated sources, peak age, or timely
throughput~\cite{tripathi2022optimizing,ramakanth2023monitoring,lin2023scheduling,fountoulakis2023scheduling}.
These problems are closest to the source-scheduling part of our decoupled
model; our action additionally selects a channel whose service rate is
unknown. Another line of work studies broadcast channels and multi-sensor systems,
together with remote-estimation and index-based scheduling
formulations~\cite{kadota2018scheduling,sun2020sampling,chang2023lightweight,xie2023scheduling,tripathi2024whittle}.
The instances in which selecting a source also exposes its source-dependent
channel or dynamics are more directly related to our coupled model. These studies assume that the relevant channel or system statistics are
known or time-invariant, whereas we estimate service rates that change
during operation.

\textit{AoI bandits and learning-based scheduling.}
When channel statistics are unknown but fixed, single-source AoI scheduling
can be formulated as a multi-armed bandit problem. The formulation
in~\cite{bhandari2020age} compares the learner with the fixed channel having
the largest service rate,
which is the AoI-optimal stationary choice under time-invariant channel service rates.
Order-optimal AoI regret for this comparator was subsequently obtained using
upper-confidence-bound and Thompson-sampling schedulers~\cite{fatale2021regret}.
Multi-source extensions include aging and decentralized bandits, while other
learning-based formulations address unknown queue service rates, HARQ, side
information, and transmission
costs~\cite{atay2021aging,prasad2021decentralized,krishnasamy2019learning,ceran2021reinforcement,stahlbuhk2019learning,chen2022age,huang2024aoi,liu2025learning}.
The stationary bandit guarantees most directly comparable to ours estimate
time-invariant service rates or transition models. Our variation-budget setting instead permits
the channel with the largest service rate to change throughout the horizon.

\textit{Non-stationary bandits and wireless scheduling.}
For time-varying channels, the total-variation model for drifting bandit rewards and the REXP3 policy were introduced in~\cite{besbes2014stochastic}. Subsequent
studies developed sliding-window methods, contextual extensions, and
variation-adaptive
wrappers~\cite{cheung2019learning,luo2018efficient,garivier2011upper,wei2021nonstationary}.
Wireless scheduling and queueing under time-varying statistics have also been
studied~\cite{huang2024lyapunov,nguyen2023learning}, but their guarantees
concern reward, throughput, or queue stability rather than the future AoI cost
of a missed update. More closely related to our setting, nonstationary AoI
bandits have been analyzed under a switching budget~\cite{song2022regret},
with a related application to asynchronous federated
learning~\cite{li2025mab}. A switching budget counts how often the service-rate vector
changes and charges a small and a large jump equally. A variation budget
instead accumulates the magnitudes of those jumps, so it also captures gradual
drift.
Complementary online AoI formulations place time variation in different
parts of the model. Adversarial channel states have been studied with
competitive guarantees~\cite{banerjee2020fundamental}. Other work varies the
AoI cost functions and compares with fixed and dynamic policy
benchmarks~\cite{tripathi2021online}. Unlike these formulations, we study AoI
scheduling with unknown, time-varying channel service rates constrained by a
total-variation budget and derive AoI-regret guarantees that account for the
future cost of missed updates.

\textit{Optimistic reinforcement learning and restless bandits.}
The coupled model is also related to model-based optimistic reinforcement learning,
which constructs a confidence set of plausible MDPs and plans in a favorable
one. This approach yields strong regret guarantees for finite-state MDPs and
has been extended to time-varying transition models using recent-data
windows~\cite{jaksch2010near,fruit2018efficient,cheung2020reinforcement,gajane2018sliding}.
A separate restless-bandit line of work assigns each source a scalar priority index to
approximate the joint scheduling problem. Such methods have been studied with
drifting parameters and, in AoI systems, with energy harvesting, channel
probing, and time-varying
dynamics~\cite{hung2025nonstationary,jaiswal2025age,shisher2025online}.
These index-based methods reduce the joint scheduling problem to a separate
score for each source. Our approach instead learns the \(M\) channel service
rates and plans over the joint state of all sources in the capped process,
while evaluating regret against the original finite-horizon oracle.

\textit{Relation to the conference version.}
A preliminary version of this work appeared in~\cite{wang2024regret}. Relative
to that version, the present paper adds channel-to-AoI regret lower bound transfers for any causal policy, together with upper bound transfers valid for Max-Age source scheduling policies. Moreover, the paper studies the
variation-budget-free Bandit over Bandit (BoB) extensions, and analyzes the AoI regret against the finite-horizon oracle in the coupled multi-source system.

\subsection{Our Contributions}

The contributions of this paper are as follows.

\begin{itemize}
\item For the single-source model, we establish an order-level relation
between AoI regret and channel reward regret in a drifting environment. This
relation applies to every causal policy, yields an AoI-regret lower bound, and
converts the channel-regret guarantee of REXP3 into an AoI-regret upper bound
that is separated from the lower bound only by logarithmic factors.

\item For the decoupled multi-source model, we separate source scheduling from
channel learning, and extend the single-source regret relation to multiple sources.
The resulting bounds explicitly show how AoI regret depends on the number of
sources, and Max-Age~+~REXP3
matches the lower bound in its time and variation dependence up to logarithmic
factors for a fixed number of sources.

\item For the coupled multi-source model, we develop an optimistic scheduler
for the MDP formed by the source ages and their associated channels. We analyze
the learner on a finite capped age space while keeping the original
finite-horizon oracle as the benchmark. The resulting bound has the same
time and variation dependence as in the single-source and decoupled
multi-source models.

\item We adapt the BoB framework to the single-source and decoupled multi-source models so
that the restart window can be learned without knowing the variation budget.
An outer EXP3 learner selects among dyadic candidate windows using group
success fractions as feedback, yielding variation-budget-free AoI-regret
bounds.

\end{itemize}

The remainder of the paper is organized as follows.
Section~\ref{sc:model} presents the system model and regret formulation.
Sections~\ref{sc:single} and~\ref{sc:decoupled} study the single-source and
decoupled multi-source models, respectively. Section~\ref{sc:bob} develops
the variation-budget-free BoB extensions, and Section~\ref{sc:coupled}
studies the coupled multi-source model. Section~\ref{sc:simulation}
reports numerical results, and Section~\ref{sc:conclusion} concludes the
paper.

\section{System Model}
\label{sc:model}

Let \(M\) and \(K\) denote the numbers of information sources and channels,
respectively, and write \([n]:=\{1,\ldots,n\}\). We study a \(T\)-slot system
in which the sources and channels are indexed by \([M]\) and \([K]\), and at
most one fresh-packet transmission is attempted per slot. We take \(K\ge2\)
and, in multi-source systems, \(M\ge2\). In the
single-source model, \(M=1\) and the scheduler selects one of the \(K\)
channels. In the decoupled multi-source model, the scheduler selects a source and a channel
separately. In the coupled multi-source model, each source has its own channel, so selecting
a source also selects its channel and \(K=M\). Table~\ref{tab:model-summary}
summarizes the three models. We display the dependence on \(K\) and \(M\)
explicitly.

Each channel is modeled as an ON/OFF channel: a transmission succeeds in the
ON state and fails in the OFF state. Let \(\mu_k(t)\) denote the service rate
of channel \(k\) in slot \(t\), namely, its probability of being ON. These
rates may change over time and are unknown to the scheduler; their values do
not react to its actions or transmission results.
In each slot, the scheduler observes the current ages, fresh packets are
generated, an action is selected, the transmission outcome is observed, and
the ages are then updated for the next slot. Let \(X_t\in\{0,1\}\) denote the transmission outcome. If channel \(k\) is selected, then
\(X_t\) is independently drawn from
\(\operatorname{Bernoulli}(\mu_k(t))\), and the scheduler observes only
\(X_t\) as channel feedback.

\subsection{AoI Dynamics and Regret}
We use beginning-of-slot AoI. Let \(A_m^\pi(t)\) be the age of source \(m\)
under policy \(\pi\), with \(A_m^\pi(1)=1\), and let \(m_t\) denote the source
selected in slot \(t\) (\(m_t=1\) in the single-source model). A fresh packet
delivered in slot \(t\) is one slot old at the beginning of slot \(t+1\),
which explains the minimum age of one:
\[
A_m^\pi(t+1)=
\begin{cases}
1, & \text{if \(m=m_t\) and \(X_t=1\),}\\
A_m^\pi(t)+1, & \text{otherwise.}
\end{cases}
\]
The equally weighted sum-AoI in slot \(t\) is
\(\sum_{m=1}^M A_m^\pi(t)\).
Let \(\pi^\star\) denote the corresponding oracle in
Table~\ref{tab:model-summary}. Our goal is to minimize the expected cumulative
sum-AoI, or equivalently its difference from the oracle, namely the AoI regret:
\begin{equation}
\label{eq:aoi_regret}
R^\pi(T)
:=
\mathbb E^\pi\!\left[
    \sum_{t=1}^T\sum_{m=1}^M A_m^\pi(t)
\right]
-
\mathbb E^{\pi^\star}\!\left[
    \sum_{t=1}^T\sum_{m=1}^M A_m^{\pi^\star}(t)
\right].
\end{equation}
For the single-source model, set \(M=1\). The expectation is over transmission
outcomes and any policy randomization. 

\begin{table*}[!t]
\centering
\caption{The three source--channel models.}
\label{tab:model-summary}
\footnotesize
\renewcommand{\arraystretch}{1.15}
\begin{adjustbox}{width=\textwidth}
\begin{tabular}{|p{0.12\textwidth}|p{0.18\textwidth}|p{0.25\textwidth}|p{0.15\textwidth}|p{0.25\textwidth}|}
\hline
\textbf{Source setting} & \textbf{Source--channel model} & \textbf{Slot-\(t\) action} & \textbf{Channel service rate} & \textbf{Oracle} \\ \hline
Single source & Single source and multi-channel & Channel \(k_t\in[K]\), with \(M=1\) & \(\mu_{k_t}(t)\) & Channel with the largest \(\mu_k(t)\) \\ \hline
\multirow{2}{*}{Multiple sources} & Decoupled multi-source and multi-channel & Pair \((m_t,k_t)\in[M]\times[K]\) & \(\mu_{k_t}(t)\) & Max-Age source + channel with the largest \(\mu_k(t)\) \\ \cline{2-5}
& Coupled multi-source and multi-channel & Source \(I_t\in[M]\), which also selects channel \(I_t\); \(K=M\) & \(\mu_{I_t}(t)\) & MDP optimum jointly accounting for source AoI and channel service rate \\ \hline
\end{tabular}
\end{adjustbox}
\end{table*}

\subsection{Channel Assumptions}
The scheduler estimates the channel service rates from past transmission
successes and failures.

\begin{assumption}[Drifting service rates]
\label{ass:drifting}
\[
\sum_{t=1}^{T-1}
\max_{k\in[K]}|\mu_k(t+1)-\mu_k(t)|\le V_T,
\]
where \(V_T=\Theta(T^\alpha)\) for some
\(0<\alpha<1\).
\end{assumption}
This assumption allows each channel service rate to change over time while
limiting the total change. The exponent \(\alpha\) specifies how the variation budget
scales with \(T\). For example, when a vehicle moves through areas with gradually changing shadowing
and interference, the success probabilities may vary in every slot but their
total movement over a trip remains limited.

\begin{assumption}[Bounded service rates]
\label{ass:bounded_rates}
For every channel \(k\) and every slot \(t\),
\[
\mu_k(t)\in[p,P],\qquad 0<p<P<1.
\]
\end{assumption}
In practical terms, every channel has a positive probability of success in
every slot. For example, a sensor that remains within base-station coverage
retains a positive chance of delivery as fading and interference change.

All subsequent results are established under
Assumptions~\ref{ass:drifting}--\ref{ass:bounded_rates}.

The following remark defines the coupled channel states used to compare
different channel choices.

\begin{remark}[Coupled Channel States]
\label{rem:coupled_channel_states}
When comparing two policies, we may use the same i.i.d.\
\(U_t\sim\operatorname{Unif}[0,1]\) in slot \(t\) and declare a transmission
over channel \(k\) successful when \(U_t\le\mu_k(t)\). Each transmission then
has the correct Bernoulli distribution, and success on a channel with a
smaller service rate implies success on one with a larger service rate. This
joint construction defines the coupled channel states used in AoI-bandit
analysis~\cite{atay2021aging}.
\end{remark}

\subsection{Problem Formulation}
The scheduler knows \(T,M,K\) and the action structure in
Table~\ref{tab:model-summary}, but not the channel service rates. It chooses
causally from the current ages and past transmission results. We study its
AoI regret. Whether
\(V_T\), \(p\), or \(P\) is known is algorithm-specific and is stated with each
algorithm.

\subsection{Notations}
We use $O(\cdot)$, $\Omega(\cdot)$, and $\Theta(\cdot)$ to describe the asymptotic growth of a function. Specifically, $O(g(n))$ gives an upper bound, meaning that the function grows no faster than $g(n)$; $\Omega(g(n))$ gives a lower bound, meaning that it grows no slower than $g(n)$; and $\Theta(g(n))$ applies if both $O(g(n))$ and $\Omega(g(n))$ hold, meaning that it grows at the same asymptotic rate as $g(n)$, up to constant factors.
We use $\widetilde{O}(\cdot)$ to suppress polylogarithmic factors. We write
\(x^+:=\max\{x,0\}\),
and \(\mathbb{I}\{\cdot\}\) denotes an indicator. 

\section{Single-Source AoI Bandits}
\label{sc:single}

In the single-source model, the action in slot \(t\) is a channel
\(k_t\in[K]\), and the transmission succeeds with probability
\(\mu_{k_t}(t)\). The learner estimates \(\mu_k(t)\) from past successful and
failed transmissions. An oracle knows every \(\mu_k(t)\) and selects a channel
with the largest value in each slot. We write
\(A^\pi(t):=A_1^\pi(t)\) for the single source's age.

\begin{figure}[htbp]
\centerline{\includegraphics[width=\modeldiagramwidth]{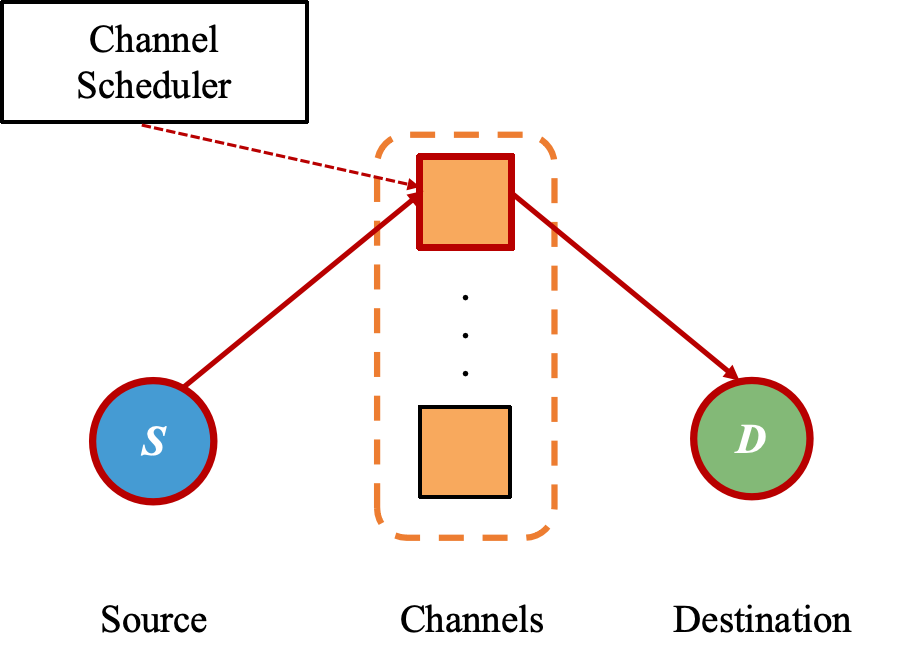}}
\caption{Single-source system. A successful transmission resets the source
age; otherwise, the age increments.}
\label{fig:single-source}
\end{figure}

Figure~\ref{fig:single-source} illustrates the single-source setting, where the scheduler
selects one of $K$ channels to transmit the source update to the
destination. We study how channel-selection errors accumulate in AoI
and how the unknown service rates can be learned.

Let \(k_t^\pi\) be the channel selected by policy \(\pi\). The oracle selects any
\(k_t^\star\in\argmax_{k\in[K]}\mu_k(t)\). This is the
single-source oracle in~\eqref{eq:aoi_regret}: a larger \(\mu_k(t)\)
cannot increase the age, so no other causal channel choice can have smaller
expected AoI. We use \(R_{\rm ch}^\pi(T)\) to denote channel regret, the
expected cumulative difference between the largest service rate and that of
the selected channel:
\[
R_{\rm ch}^\pi(T)
:=
\sum_{t=1}^T
\mathbb E\!\left[
\max_{k\in[K]}\mu_k(t)-\mu_{k_t^\pi}(t)
\right].
\]
It counts the loss in the current slot, whereas AoI regret also includes the
later age cost caused by a missed update.

\subsection{Lower Bound}

A channel-regret lower bound does not directly imply an AoI regret lower
bound. Channel regret measures only the loss in the current slot,
whereas if the oracle succeeds and the learner fails, the learner
retains its previous age instead of resetting it. If subsequent
transmissions also fail, this larger age continues to contribute to
the cumulative AoI. For a lower bound, however, it is sufficient to show that these channel-selection errors create a comparable AoI regret in the first affected slot.
In Lemma \ref{lem:single_lower_transfer}, \(g(T)\) denotes a function of \(T\).

\begin{lemma}
\label{lem:single_lower_transfer}
If a causal policy for the single-source system has channel regret
\(\Omega(g(T))\), then its AoI regret is \(\Omega(g(T))\).
\end{lemma}

Under the coupling in Remark~\ref{rem:coupled_channel_states}, every learner success is also
an oracle success. Hence an oracle-only success in slot \(t\) creates an AoI gap
in slot \(t+1\). Summing these next-slot gaps gives the channel regret up to a
constant horizon-boundary term. Therefore, a channel-regret lower bound
transfers to an AoI-regret lower bound without changing its order. See
Supplementary Appendix~\ref{app:single_lower_transfer} for proof details.

Lemma~\ref{lem:single_lower_transfer} reduces the AoI regret lower-bound
problem to constructing a hard drifting channel sequence. We use the standard
non-stationary bandit construction and obtain Theorem \ref{th:single-lower}.

\begin{theorem}[Single-source system lower bound]
\label{th:single-lower}
Every causal policy has AoI regret
\(\Omega((KV_T)^{1/3}T^{2/3})\) for some channel service-rate sequence in the single-source system.
\end{theorem}

The construction partitions the horizon into blocks and changes the
best channel across blocks, so observations collected in one block
do not identify the best channel in the next. After shifting and
scaling the Bernoulli means into \([p,P]\), the standard
drifting-bandit lower bound applies, and
Lemma~\ref{lem:single_lower_transfer} transfers the resulting channel regret
to AoI regret. This result shows that no causal scheduler can guarantee an
AoI-regret order smaller than \((KV_T)^{1/3}T^{2/3}\) over the considered class
of service-rate sequences.

\subsection{REXP3 Upper Bound}

For the upper bound, the stationary argument
in~\cite{fatale2021regret} does not directly extend to our setting. That
argument reorders the uses of suboptimal channels relative to a fixed best
channel, whereas the best channel here may change from slot to slot. The remaining question is whether a missed update can create a persistent AoI gap even when channel regret is small. The next lemma shows that Assumption 2 limits the resulting amplification to a logarithmic factor.

\begin{lemma}
\label{lem:single_upper_transfer}
In the single-source system, if a causal policy has channel regret \(O(g(T))\),
then its AoI regret is \(O(\log T\,g(T))\).
\end{lemma}

Under Remark~\ref{rem:coupled_channel_states}, an oracle-only success creates
an AoI gap. The next learner success is also an oracle success, so both ages
reset and this gap is removed. Assumption~\ref{ass:bounded_rates} guarantees that every learner
transmission succeeds with probability at least \(p\), which implies that the
probability of a long interval without a learner success decays exponentially
with its length. Truncating such intervals at a logarithmic threshold makes
the remaining tail contribution negligible, while the contribution below the
threshold introduces only a logarithmic factor. Thus
Lemma~\ref{lem:single_upper_transfer} converts a channel-regret upper bound
into an AoI-regret upper bound with only logarithmic loss. See Supplementary
Appendix~\ref{app:single_upper_transfer} for proof details.

It therefore remains to control channel regret under drifting service rates. REXP3 does so by periodically restarting EXP3 and discarding observations that may no longer describe the current channels, leading to Theorem \ref{th:rexp3}.

\begin{algorithm}[!t]
\caption{Restarted EXP3 scheduler}
\label{alg:rexp3-channel}
\algorithmtextformat
\begin{algorithmic}[1]
\Statex \textbf{Input:} Horizon \(T\), channels \(K\), and variation budget
\(V_T\).
\Statex \textbf{Output:} Channel decisions \(k_1,\ldots,k_T\).
\State Set
\(L_{\rm rst}\gets\min\{T,\lceil(K\log K)^{1/3}(T/V_T)^{2/3}\rceil\}\).
\State Set
\(\gamma\gets\min\{1,\sqrt{K\log K/((e-1)L_{\rm rst})}\}\).
\State Partition \([T]\) into restart segments of length \(L_{\rm rst}\), truncating the final segment if needed.
\For{each restart segment \(B\)}
    \State Initialize \(w_k\gets1\), \(k\in[K]\).
    \For{each slot \(t\in B\)}
        \State Compute \(q_{k,t}\gets (1-\gamma)w_k/\sum_{\ell=1}^K w_\ell+\gamma/K\), \(k\in[K]\).
        \State Sample \(k_t\sim q_{\cdot,t}\) and transmit over \(k_t\).
        \State Form \(\widehat X_{k,t}\gets X_t\mathbb{I}\{k=k_t\}/q_{k,t}\), \(k\in[K]\).
        \State Update \(w_k\gets w_k\exp((\gamma/K)\widehat X_{k,t})\), \(k\in[K]\).
    \EndFor
\EndFor
\end{algorithmic}
\end{algorithm}

\begin{theorem}[REXP3 upper bound]
\label{th:rexp3}
Algorithm~\ref{alg:rexp3-channel} has AoI regret
\(\widetilde O((KV_T)^{1/3}T^{2/3})\).
\end{theorem}

The restart length balances two sources of channel regret. Short
restart segments increase the repeated learning cost, whereas long
segments increase the error caused by service-rate variation within
each segment. Balancing these terms gives the restart length in
Algorithm~\ref{alg:rexp3-channel} and the corresponding REXP3
channel-regret bound. Lemma~\ref{lem:single_upper_transfer} then converts this
channel bound to AoI regret. Together with
Theorem~\ref{th:single-lower}, the result shows that the single-source lower
and upper bounds have the same polynomial dependence on \(K\), \(V_T\), and
\(T\), up to logarithmic factors.

\section{Decoupled Multi-Source AoI Bandits}
\label{sc:decoupled}

\begin{figure}[htbp]
\centerline{\includegraphics[width=\modeldiagramwidth]{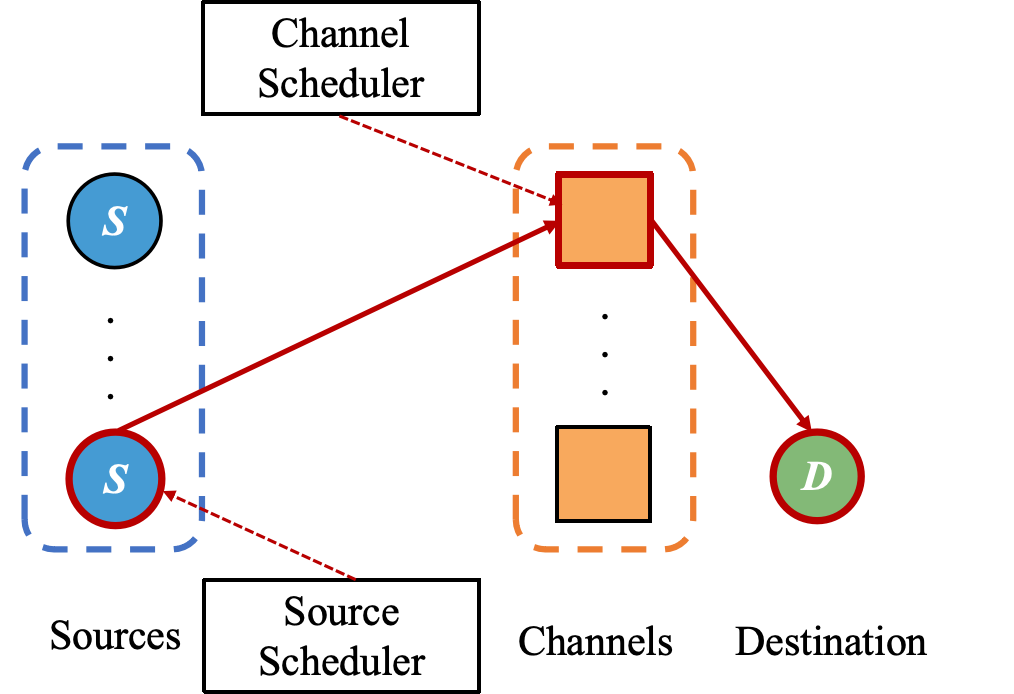}}
\caption{Decoupled multi-source system. Each slot selects a source and a channel separately; the
channel determines whether delivery succeeds, and the source determines which
age can reset.}
\label{fig:decoupled}
\end{figure}

In the decoupled model, the action in slot \(t\) is a source--channel pair
\((m_t,k_t)\in[M]\times[K]\), and the transmission succeeds with
probability \(\mu_{k_t}(t)\). Figure~\ref{fig:decoupled} illustrates this
action. The source decision determines which age resets after a success, and
the channel decision determines whether the transmission succeeds. Because
\(\mu_{k_t}(t)\) does not depend on \(m_t\), changing the source does not
change whether the transmission succeeds.

We first derive channel-to-AoI regret transfers for the decoupled
model and then combine them with Max-Age source scheduling and
REXP3 channel learning. Compared with the single-source model, one
missed success can affect several source ages and remain visible in
subsequent slots. The lower and upper transfers below account for
this additional persistence.

\subsection{Lower Bound}

The one-slot comparison underlying Lemma 1 remains valid with multiple sources, but it ignores most of the resulting AoI loss. A missed success can leave several source ages stale and remain visible over multiple subsequent slots. The next lemma quantifies this persistence and shows that it amplifies a channel-regret lower bound by a quadratic factor in $M$.

\begin{lemma}
\label{lem:decoupled_lower_transfer}
For the decoupled multi-source system with \(M\) sources, any causal
policy \(\pi\) with channel regret \(\Omega(g(T))\) incurs AoI regret
\(\Omega(M^2g(T))\).
\end{lemma}

To establish this result, we first replace the learner's source schedule by Max-Age while
keeping the same success--failure sequence;
Proposition~\ref{prop:decoupled_separation} shows that this replacement cannot
increase the AoI. We then add the oracle-only successes in reverse
chronological order. Each added success affects the next \(M\) Max-Age states
and contributes at least \(M,M-1,\ldots,1\) to the corresponding reductions
in sum-AoI, apart from horizon-boundary terms. The resulting cumulative
contribution is of order \(M^2\). Thus
Lemma~\ref{lem:decoupled_lower_transfer} shows that, in the decoupled model,
the AoI penalty associated with channel regret scales quadratically with the
number of sources. See Supplementary
Appendix~\ref{app:decoupled_lower_transfer} for proof details.


Applying Lemma \ref{lem:decoupled_lower_transfer} to the hard drifting channel sequence used for Theorem \ref{th:single-lower} gives the decoupled multi-source system AoI regret lower bound in Theorem~ \ref{th:decoupled_lb}.

\begin{theorem}[Decoupled multi-source system lower bound]
\label{th:decoupled_lb}
Every causal decoupled policy has AoI regret
\(\Omega(M^2(KV_T)^{1/3}T^{2/3})\) for some channel service-rate sequence in the decoupled multi-source system.
\end{theorem}

The dependence on \(K\), \(V_T\), and \(T\) remains the same as in the single-source lower bound, while the additional \(M^2\) factor accounts for the persistence of a missed success across multiple source ages.

\subsection{Max-Age + REXP3 Upper Bound}

After separating channel learning, the source decision remains
state dependent because a successful transmission resets only the
selected source. To separate source scheduling from channel
selection, we need a source rule that is optimal for every fixed
success--failure sequence. Proposition \ref{prop:decoupled_separation} shows that choosing the source with the maximum age minimizes AoI for the decoupled multi-source system.

\begin{proposition}[Max-Age reduction]
\label{prop:decoupled_separation}
Under every fixed success--failure sequence, Max-Age minimizes cumulative
sum-AoI pathwise. The oracle in the decoupled multi-source system therefore serves a Max-Age source and
selects a channel with the largest service rate.
\end{proposition}

The proof sorts the age vectors in decreasing order and compares the
corresponding entries after every slot. A failure increments all
ages and preserves this order. After a success, resetting the largest
age produces a sorted next vector whose entries are no larger than
the corresponding entries obtained by resetting any other source.
Induction therefore gives pathwise optimality of Max-Age. This
property allows source scheduling and channel selection to be
analyzed separately. See Supplementary
Appendix~\ref{app:decoupled_max_age} for proof details.

For an upper bound, it is not sufficient to bound only the
immediate effect of an oracle-only success. We must control both the
magnitude of the resulting AoI gap and the number of slots for which
the gap persists. Let
\(\Pi_{\rm MA}\) denote the causal decoupled policies that select a source
with the largest current age in every slot.

\begin{lemma}
\label{lem:decoupled_upper_transfer}
For the decoupled multi-source system with $M$ sources, if
\(\pi\in\Pi_{\rm MA}\) has channel regret \(O(g(T))\), then its AoI regret is
\(O(M(M+\log T)g(T))\).
\end{lemma}

Representing a Max-Age state by its \(M\) most recent successful-update times makes these two effects explicit. For each oracle-only success, the gap magnitude is controlled by the
distance to the oracle's \(M\)-th most recent success. The gap is
removed after the learner obtains \(M\) subsequent successes, which
replace the relevant recent-update times in the Max-Age state.
Assumption~\ref{ass:bounded_rates} gives an exponentially decaying tail for the first
quantity and a bounded expected clearing time for the second.
Combining these bounds yields the factor in
Lemma~\ref{lem:decoupled_upper_transfer}. Thus the lemma shows that the
persistent multi-source effect of channel errors remains controlled by
channel regret. See Supplementary
Appendix~\ref{app:decoupled_upper_transfer} for proof details.

\begin{algorithm}[!t]
\caption{Max-Age + REXP3 scheduler}
\label{alg:max-age-rexp3}
\algorithmtextformat
\begin{algorithmic}[1]
\Statex \textbf{Input:} \(T,M,K,V_T\). \textbf{Output:} \((m_t,k_t)_{t=1}^T\).
\State Initialize \(A_m(1)\gets1\), \(m\in[M]\), and REXP3 as in Algorithm~\ref{alg:rexp3-channel}.
\For{each REXP3 restart segment \(B\)}
    \State Reset the REXP3 channel weights.
    \For{each slot \(t\in B\)}
        \State Select the smallest index \(m_t\in\argmax_{m\in[M]}A_m(t)\).
        \State Execute the next REXP3 channel decision, transmit source \(m_t\), and observe \(X_t\).
        \State Update REXP3 and the ages according to Section~\ref{sc:model}.
    \EndFor
\EndFor
\end{algorithmic}
\end{algorithm}

REXP3 controls the channel-learning error, while
Lemma~\ref{lem:decoupled_upper_transfer} accounts for the cumulative AoI effect
of these errors across multiple sources. Combining the two gives Theorem \ref{th:decoupled_ub}. 

\begin{theorem}[Max-Age + REXP3 upper bound]
\label{th:decoupled_ub}
Algorithm~\ref{alg:max-age-rexp3} has AoI regret
\(\widetilde O(M^2(KV_T)^{1/3}T^{2/3})\).
\end{theorem}

Together with Theorem~\ref{th:decoupled_lb}, the result shows that the decoupled lower
and upper bounds have the same polynomial dependence on \(M\), \(K\), \(V_T\),
and \(T\), up to logarithmic factors.

\section{Variation-Budget-Free Channel Scheduling}
\label{sc:bob}

When \(V_T\) is unknown, Algorithm~\ref{alg:rexp3-channel} cannot use a
restart window tuned to the variation budget.
BoB~\cite{cheung2019learning} learns the window online, but group AoI is not
suitable as the outer-bandit feedback because it depends on the age inherited
at the beginning of the group, which was generated by earlier window choices.
We therefore use the observed success fraction in the current group.

\begin{figure}[!t]
\centerline{\includegraphics[width=\widediagramwidth]{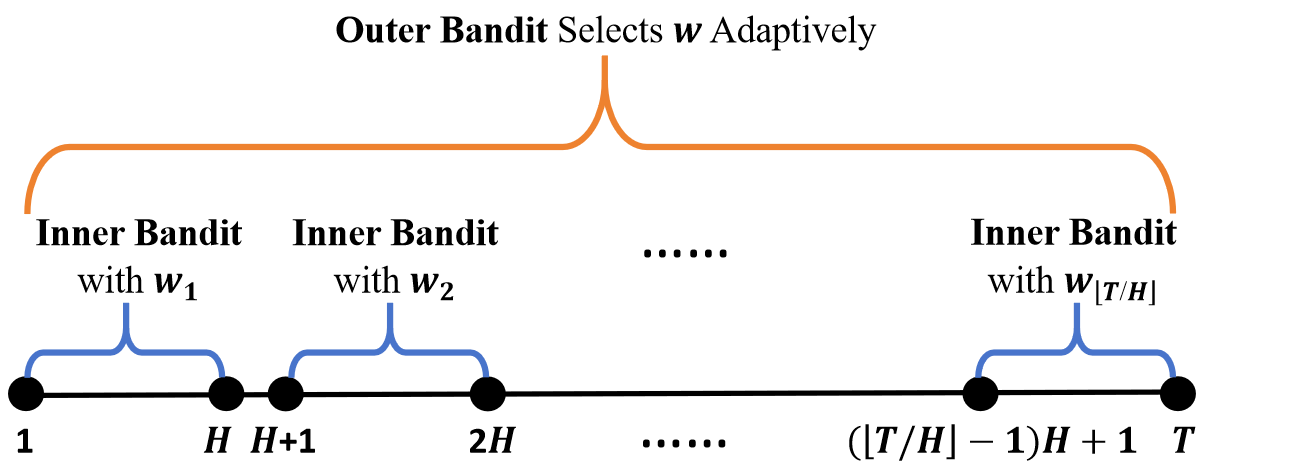}}
\caption{BoB chooses one REXP3 restart window per group. The observed success
fraction in that group is fed back to outer EXP3.}
\label{fig:bob-wrapper}
\end{figure}

Figure~\ref{fig:bob-wrapper} illustrates the BoB wrapper. It divides the horizon
into groups and uses an outer EXP3 instance to select one
restart window for each group. Given the selected window
\(w\), the wrapper runs an inner AoI scheduler \(\pi_w\)
throughout that group.

The inner scheduler can be changed without modifying the
outer wrapper. In the single-source model, \(\pi_w\) is REXP3
with restart window \(w\); in the decoupled model, it is
Max-Age~+~REXP3 with the same restart window. Only the
learning state of the inner channel learner is restarted. The
actual AoI continues across all restart segments and groups.

Algorithm~\ref{alg:bob} uses groups of length \(G_B\) and the
logarithmic-size dyadic family
\(\mathcal W:=\{2^j:0\le j\le\lfloor\log_2G_B\rfloor\}\cup\{G_B\}\).
The group scale is chosen so that the cost of learning the restart
window and the worst-case cost of restricting the window to one
group have the same polynomial order. The dyadic family keeps
only logarithmically many outer arms while containing a
constant-factor approximation to every relevant restart scale.
This grid is convenient for the analysis rather than essential
to the implementation; any polylogarithmic-size candidate
family with comparable coverage can be used.

\begin{algorithm}[t]
\caption{BoB wrapper for AoI bandits}
\label{alg:bob}
\algorithmtextformat
\begin{algorithmic}[1]
\Require Horizon \(T\), channels \(K\), candidate windows
\(\mathcal W\), and inner AoI scheduler \(\pi_w\).
\State Set \(G_B\leftarrow \min\{T,\lceil\sqrt{KT}\rceil\}\) and
initialize outer EXP3 over \(\mathcal W\).
\For{each group \(\mathcal I\) of at most \(G_B\) slots}
    \State Select \(w\in\mathcal W\) using outer EXP3.
    \State Run \(\pi_w\) in \(\mathcal I\), restarting its
    learning state every \(w\) slots while keeping the actual ages.
    \State Feed the group success fraction back to outer EXP3.
\EndFor
\end{algorithmic}
\end{algorithm}

\begin{corollary}[Single-source system BoB upper bound]
\label{cor:single-bob}
In the single-source system, the REXP3-BoB policy
has AoI regret \(\widetilde O\!\left((KV_T)^{1/3}T^{2/3}+K^{1/4}T^{3/4}\right)\).
\end{corollary}

The success fraction is bounded in \((0,1)\) and depends only on
transmission outcomes within the current group. It therefore provides
valid feedback for outer EXP3 without carrying the starting age from
previous groups into the comparison. Combining the outer learning
cost with the best fixed-window REXP3 bound, and then applying
Lemma~\ref{lem:single_upper_transfer}, gives the corollary. The additional
\(K^{1/4}T^{3/4}\) term is the cost of adapting the restart scale without
knowing \(V_T\). See Supplementary Appendix~\ref{app:single_bob_proof} for proof details.

In the decoupled multi-source system, Max-Age does not change the BoB
channel-learning problem because source selection does not affect the
transmission success probability.

\begin{corollary}[Decoupled multi-source system BoB upper bound]
\label{cor:decoupled-bob}
In the decoupled multi-source system, Max-Age with REXP3-BoB has AoI regret \(\widetilde O\!\left(M^2\bigl[(KV_T)^{1/3}T^{2/3}+K^{1/4}T^{3/4}\bigr]\right)\).
\end{corollary}

The BoB channel-regret analysis is therefore unchanged from
Corollary~\ref{cor:single-bob}. Applying the multi-source system upper bound in
Lemma~\ref{lem:decoupled_upper_transfer} gives the stated AoI bound. See Supplementary Appendix~\ref{app:decoupled_bob_proof} for proof details.
\FloatBarrier
\section{Coupled Source-Channel AoI MDP}
\label{sc:coupled}

In the coupled model, source \(i\) is tied to channel \(i\), so there are
\(K=M\) channels. The action \(I_t\in[M]\) selects source
\(I_t\) and its channel; the transmission succeeds with probability
\(\mu_{I_t}(t)\), and a success resets only the age of that source. The source and channel
decisions therefore cannot be separated. Selecting only the channel with the
largest service rate may leave an old source waiting, while selecting only the
oldest source may use a channel with a small service rate. Because each action
depends on the current ages and also changes their future values, this joint
scheduling problem can be formulated as an MDP.

\begin{figure}[htbp]
\centerline{\includegraphics[width=\modeldiagramwidth]{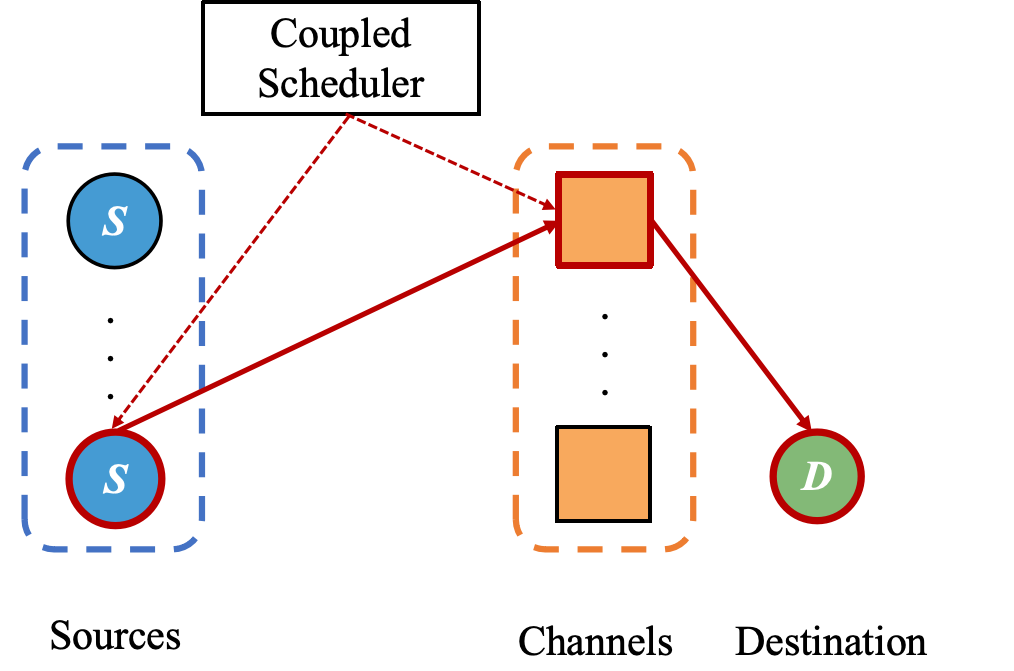}}
\caption{Coupled multi-source system. Selecting a source also selects its channel, so the
action determines both the channel service rate and the source whose age can
reset.}
\label{fig:coupled}
\end{figure}

Figure~\ref{fig:coupled} illustrates the coupled action. This section studies
how to learn the channel service rates while scheduling the sources and
comparing with the finite-horizon MDP oracle.

\vspace{-1.0\baselineskip}
\begingroup
\setlength{\abovedisplayskip}{5pt}
\setlength{\belowdisplayskip}{5pt}
\setlength{\abovedisplayshortskip}{3pt}
\setlength{\belowdisplayshortskip}{3pt}
\subsection{Optimistic Scheduler for the Capped AoI MDP Process}
\label{subsec:coupled-scheduler}

The scheduler repeats three steps: estimate the service rates from recent
transmissions, solve an optimistic MDP, and follow its policy until an
observation count roughly doubles.

Divide the horizon into blocks of at most \(L\) slots and reset the observations
at each block boundary. Each block is further divided into episodes, so every episode
lies inside one block. The
service-rate estimates and scheduling rule remain fixed in an episode, and a
new episode starts when a channel count roughly doubles. Choose an integer age cap \(Q\); the planner observes
\[
S_Q(t):=(\min\{A_1^\pi(t),Q\},\ldots,\min\{A_M^\pi(t),Q\})\in[Q]^M.
\]
The cap makes the planning state finite, while the actual ages remain uncapped.
The scheduler knows \(T,M,V_T,p,P\), but not the service rates.

At the start of episode $k$, let $N_{k,i}$ be the number of times
channel $i$ was selected earlier in the current block, and let
$\widehat{\mu}_{k,i}$ be its observed success fraction. For
$N_{k,i}>0$, define the confidence radius for channel $i$ in episode
$k$ as
\[
\beta_{k,i}
:=
C\sqrt{\frac{\log(MT/\delta)}{N_{k,i}}},
\]
where $\delta\in(0,1)$ is the confidence parameter and $C>0$ is a
sufficiently large numerical constant. This choice has the same
$N_{k,i}^{-1/2}$ scaling as the reward confidence interval used in
UCRL2~\cite[Eq.~(3)]{jaksch2010near}. The service rate used for planning is
\begin{equation}
\label{eq:main_optimistic_parameter}
q_{k,i}
=
\begin{cases}
P, & N_{k,i}=0,\\[1mm]
\min\{P,\max\{p,\widehat\mu_{k,i}+\beta_{k,i}\}\},
& N_{k,i}>0.
\end{cases}
\end{equation}
Thus \(q_k=(q_{k,1},\ldots,q_{k,M})\) is statistically optimistic for the sampled
within-block average. 

Hold \(q_k\) fixed during episode \(k\). Write a state in the capped process as
\(s=(s_1,\ldots,s_M)\in[Q]^M\). Its one-slot sum-AoI is \(\sum_m s_m\). After a
failure, the next state is \(f(s)\), where
\(f(s)_m=\min\{s_m+1,Q\}\). After a success on channel \(i\), the next state is
\(r_i(s)\): source \(i\) resets to one and every other source has the same age
as in \(f(s)\).

For fixed \(q_k\), the capped process is a finite-state average-cost MDP. Let \(h_Q(q_k,s)\) denote a bias function~\cite[Sec.~8.2.1]{puterman1994markov} satisfying its average-cost optimality equation (ACOE)~\cite[Sec.~8.4.1]{puterman1994markov}; its differences compare future capped AoI across states. The ACOE conditions and Bellman relations used below are justified in Supplementary Appendix~\ref{app:coupled} using standard finite-state average-cost MDP results~\cite[Ch.~8]{puterman1994markov}. Define the reset benefit
of source \(i\) by
\[
D_{Q,i}(q_k,s):=h_Q(q_k,f(s))-h_Q(q_k,r_i(s)).
\]
It is the future cost saved by a success rather than a failure. The ACOE then gives the Bellman selector:
\begin{equation}
\label{eq:main_canonical_selector}
I_t
=
\min\argmax_{i\in[M]}
q_{k,i}D_{Q,i}(q_k,S_Q(t)).
\end{equation}
This score accounts for both the channel service rate and the source age,
unlike a channel-only learner. For ties, we select the smallest index.

Algorithm~\ref{alg:capped_optimistic_finite_state} summarizes the procedure.
The displayed \(L,Q,\delta\) are the sufficient choices. The estimate--plan--replan loop uses standard optimistic-MDP ingredients from UCRL2~\cite{jaksch2010near}, while block restarts discard old data under drift~\cite{cheung2020reinforcement}. Our AoI-specific changes are to plan with
capped ages and to value which source a success would reset.

\begin{algorithm}
\caption{Optimistic scheduler for the capped AoI MDP process}
\label{alg:capped_optimistic_finite_state}
\algorithmtextformat
\begin{algorithmic}[1]

\Statex \textbf{Input:} horizon \(T\), number of sources \(M\),
variation budget \(V_T\), and service-rate bounds \(p,P\).

\Statex \textbf{Parameters:}
block length
\(L=\min\{T,\lceil M^{1/3}(T/V_T)^{2/3}\rceil\}\),
confidence level \(\delta=T^{-3}\), and age cap
\[
Q=
\left\lceil
C_{p,P}
\bigl(M^6p^{-2M}+Mp^{-M}(1+\log T)\bigr)
\right\rceil .
\]

\Statex \textbf{Within-block statistics:}
\(N_i\) is the number of transmissions over channel \(i\), and
\(\widehat{\mu}_i\) is its empirical service rate.
At the start of episode \(k\), \(N_{k,i}\) records the current value of \(N_i\).

\State Partition the horizon into blocks of at most \(L\) slots.

\For{each block}
    \State Initialize \(N_i\gets0\) and \(\widehat{\mu}_i\gets0\)
    for all \(i\in[M]\).

    \While{the block has not ended}
        \State Start episode \(k\) and record
        \(N_{k,i}\gets N_i\) for all \(i\in[M]\).

        \State Construct the optimistic service vector \(q_k\)
        using \eqref{eq:main_optimistic_parameter}, and solve the
        capped MDP for \(h_Q(q_k,\cdot)\).

        \While{the block has not ended and no channel count has doubled}
            \State Form \(S_Q(t)\) by capping the current actual ages at \(Q\).

            \State Select source \(I_t\) according to
            \eqref{eq:main_canonical_selector}.

            \State Transmit source \(I_t\), observe the success indicator
            \(X_t\), and update its count, empirical service rate,
            and the actual ages.
        \EndWhile
    \EndWhile
\EndFor
\end{algorithmic}
\end{algorithm}

\subsection{AoI-Regret Upper Bound}
\label{subsubsec:coupled_comparator}

The coupled model cannot use the channel-to-AoI reductions above.
Selecting a source also selects its channel and changes the future
age state, so channel learning and source scheduling must be analyzed
jointly. A direct finite-state MDP analysis is also unavailable
because the AoI state space is unbounded. Capping both the learner
and the oracle would make the state space finite but would also
change the finite-horizon benchmark. We therefore use capped ages
only for planning and analysis, while keeping the actual ages and
the original oracle uncapped.

\begin{theorem}[Coupled multi-source system optimistic scheduler upper bound]
\label{thm:coupled}
Algorithm~\ref{alg:capped_optimistic_finite_state}, using a restart length of
order \(M^{1/3}(T/V_T)^{2/3}\) and the age cap \(Q\) specified in the algorithm, has AoI regret
\(O\!\left(p^{-M}M^{7/3}Q(1+\log(MT))V_T^{1/3}T^{2/3}\right)\). The scaling with respect to $V_T$ and $T$ is
\(\widetilde O(V_T^{1/3}T^{2/3})\).
\end{theorem}

The proof combines standard optimistic-MDP learning arguments with AoI-specific structural bounds. We first establish a
uniform bound on the bias span, which measures the largest change in
future capped cost caused by changing the current state in the capped
process. This
bound converts service-rate estimation error and within-block drift
into the regret computed using ages capped at \(Q\). Although the capped MDP contains \(Q^M\)
states, the unknown transition parameters are only the \(M\)
source-specific service rates, so observations can be pooled by
source. We then derive a sharper one-coordinate bias-increment bound. Together
with a linear lower bound on the reset benefit of a source with
sufficiently large age, this shows that such a source eventually has
a larger scheduling score than recently updated sources. This
prevents persistent starvation and yields an exponentially decaying
overflow bound. The block length then balances estimation error and within-block drift. Combining the capped-regret and overflow
bounds proves Theorem~\ref{thm:coupled}. For fixed model parameters, the
resulting dependence on \(V_T\) and \(T\) agrees with that of the single-source
and decoupled multi-source models up to logarithmic factors. See Supplementary
Appendix~\ref{app:coupled} for proof details. 

\endgroup

\section{Simulation}
\label{sc:simulation}

We use simulations to check the predicted horizon scaling in the single-source
and decoupled multi-source models and to compare age-aware schedulers in the coupled model.

\subsection{Setup}

All experiments use \(K=3\) channels with service rates in \([0.1,0.9]\), and
the two multi-source models use \(M=3\) sources. In every model, the variation
grows as \(V_T=\Theta(T^\alpha)\). Model-specific horizons and service-rate
paths are given below.

\subsection{Single-Source Model}

The single-source experiment uses \(M=1\) and horizons from \(10^5\) to
\(10^7\), extended to \(3\times10^7\) for BoB. The channel with the largest
service rate changes at random block boundaries, with jumps chosen so that
\(V_T=0.22T^\alpha/K\). We test the five values of \(\alpha\) shown in the
figures and use 64 Monte Carlo replicates.

\begin{figure}[!t]
\centerline{\includegraphics[width=\simulationfigurewidth]{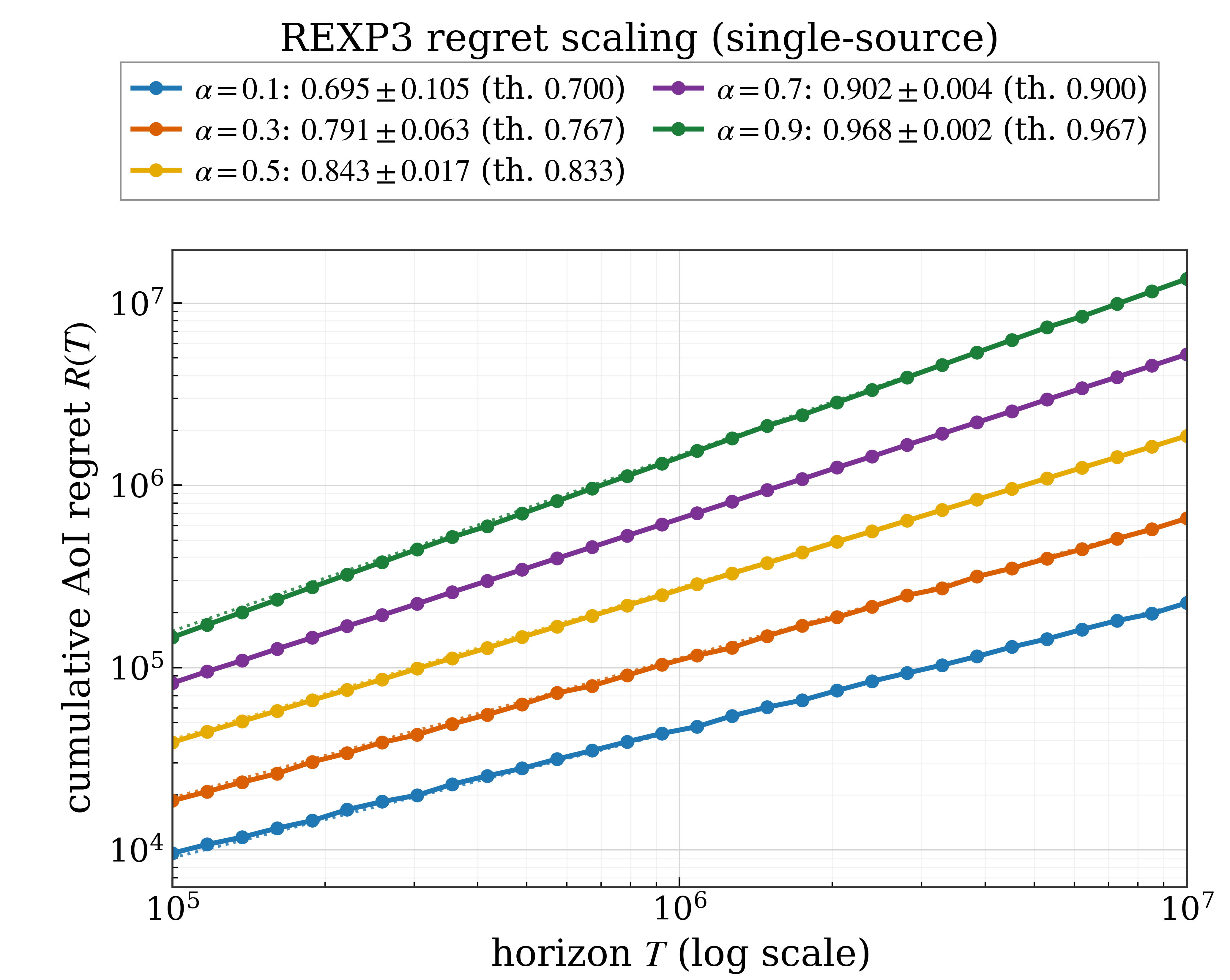}}
\caption{Single-source REXP3 regret versus horizon for five values of
\(\alpha\). The reference slope is \((2+\alpha)/3\).}
\label{fig:single-rexp3-bound}
\end{figure}

Figure~\ref{fig:single-rexp3-bound} first shows the single-source results.
Each curve corresponds to one of the five variation exponents listed in the
legend. For each seed-indexed replicate, we fit a log--log slope to the data
at the six largest values of \(T\). The legend reports the mean and one standard
deviation of the fitted slopes, while the dotted line shows the reference
exponent. Across all five
values of \(\alpha\), the fitted slopes are close to the predicted
\((2+\alpha)/3\) exponent, where $V_T^{1/3}T^{2/3}=T^{(2+\alpha)/3}$.

\begin{figure}[!t]
\centerline{\includegraphics[width=\simulationfigurewidth]{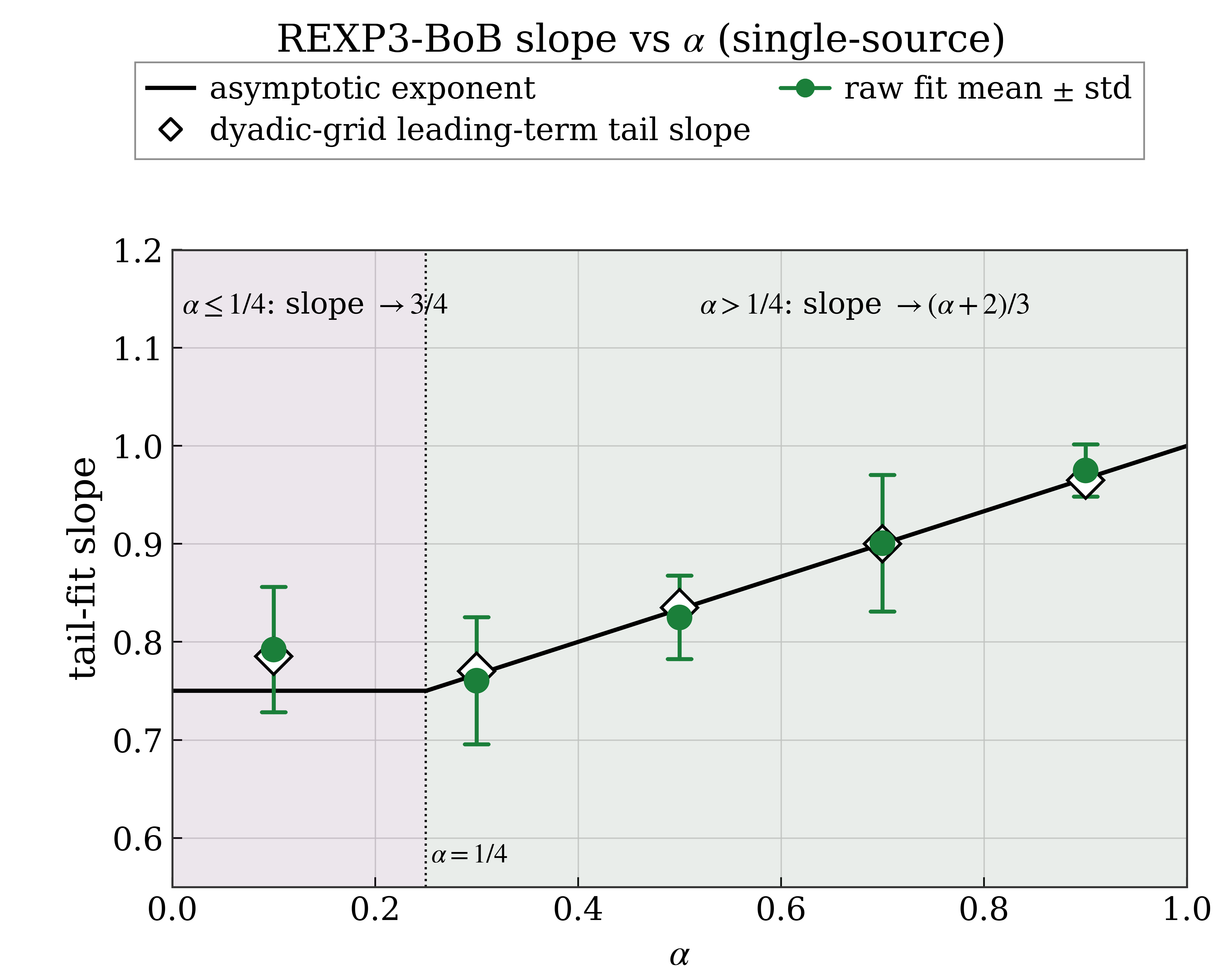}}
\caption{Single-source REXP3-BoB fitted slopes as the variation exponent
changes. The solid line shows the theoretical asymptotic exponent as time approaches infinity. The diamonds show the finite-horizon exponent. The dots show the mean fitted slopes from simulations, with error bars showing one standard deviation across Monte Carlo replicates.}
\label{fig:single-rexp3bob-slope}
\end{figure}

Figure~\ref{fig:single-rexp3bob-slope} evaluates the BoB wrapper as the variation exponent changes. 
Instead of showing five regret curves, it plots the fitted slope
against the same five values of \(\alpha\). Each green point and error bar is
the mean \(\pm\) one standard deviation across the Monte Carlo replicates.
We fit one slope per replicate using the data at the six largest values of
\(T\). The black line gives the theoretical asymptotic exponent as time approaches infinity. The diamonds show the finite-horizon fitted slopes of the dominant terms
$T^{3/4}\sqrt{J_B(T)\log J_B(T)}$ for $\alpha\le1/4$ and
$T^{(2+\alpha)/3}$ otherwise; for small $\alpha$, the extra factor $\sqrt{J_B(T)\log J_B(T)}$ makes the slope slightly larger than the asymptotic value $3/4$. Both follow the fitted
slopes across the tested variation levels.
In the simulations, we use an $O(\log T)$-size candidate set that
contains the best fixed window. The general BoB bound therefore gives
the same asymptotic order as Corollary~\ref{cor:single-bob}. The dyadic candidate set gives
similar fitted slopes.

\subsection{Decoupled Multi-Source Model}

The decoupled experiment uses \(M=3\) and otherwise follows the single-source
experiment: it uses the same horizons, service-rate paths, five values of
\(\alpha\), and 64 Monte Carlo replicates.

\begin{figure}[!t]
\centerline{\includegraphics[width=\simulationfigurewidth]{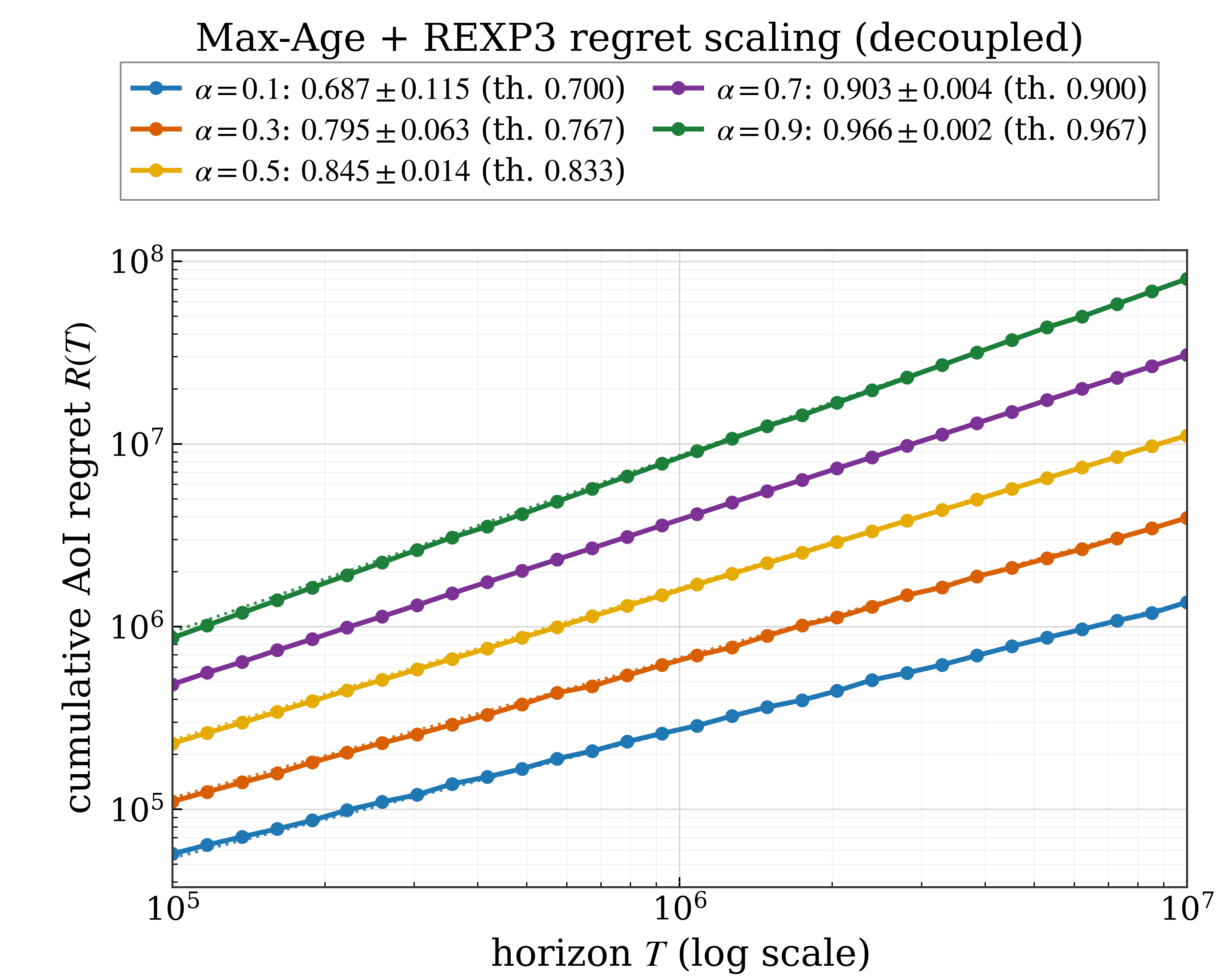}}
\caption{Decoupled Max-Age + REXP3 regret versus horizon for five values of
\(\alpha\), with reference slope \((2+\alpha)/3\).}
\label{fig:decoupled-rexp3-bound}
\end{figure}

Figure~\ref{fig:decoupled-rexp3-bound} repeats the scaling experiment for the
decoupled model. Again, each curve corresponds to one of the five values of
\(\alpha\). The legend reports the mean \(\pm\) one standard deviation of the
fitted slopes. We fit each slope to the data at the six largest values of
\(T\). The slopes are close to the reference exponent for all five variation
levels. Relative to the single-source experiment, the regret level is higher, while the
observed exponent remains essentially unchanged, consistent with
the analysis.

\begin{figure}[!t]
\centerline{\includegraphics[width=\simulationfigurewidth]{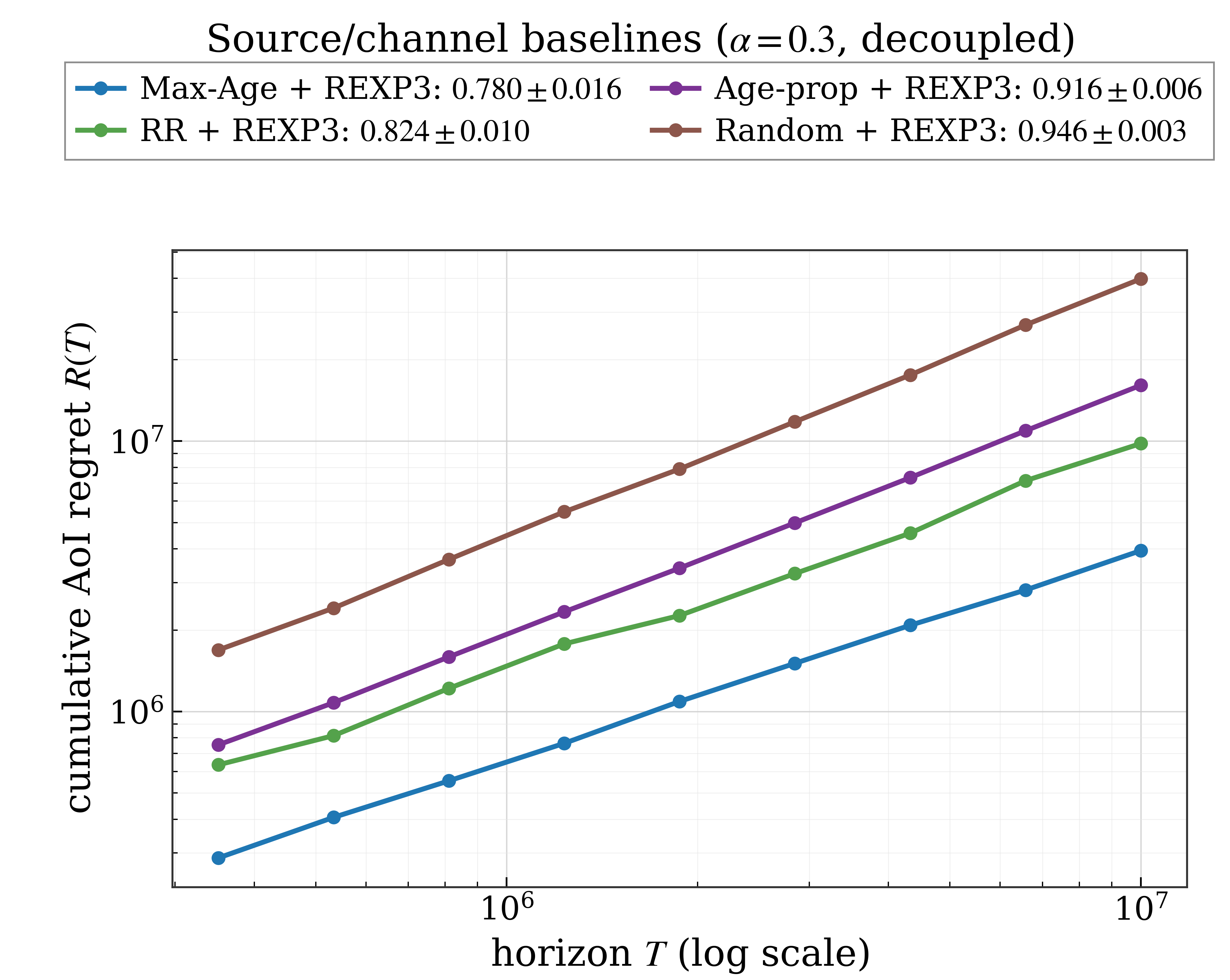}}
\caption{Decoupled source baselines at \(\alpha=0.3\).
Max-Age + REXP3 has the lowest regret level and the smallest fitted slope
among the tested ways of choosing a source.}
\label{fig:decoupled-source-baselines}
\end{figure}

Figure~\ref{fig:decoupled-source-baselines} then isolates the source decision.
Its four colored curves compare Max-Age, round-robin, age-proportional, and
random source scheduling at \(\alpha=0.3\), using nine horizons from
\(3.5\times10^5\) to \(10^7\). All four ways of choosing a source use the same REXP3
channel learner~\cite{besbes2014stochastic}. Oldest-first greedy scheduling and randomized source selection have been
studied for unreliable broadcast channels~\cite{kadota2018scheduling};
these motivate our Max-Age and Random baselines. Round-robin cycles deterministically through the sources, while
age-proportional sampling selects a source with probability proportional to
its current age. Together, these rules isolate the value of using age
systematically without changing the channel learner.
On the common Monte Carlo replicates, all four rules receive exactly the same
REXP3 channel decisions and transmission outcomes; Max-Age is also run on
additional replicates for a more stable estimate. The legend reports the
mean \(\pm\) one standard deviation of the per-replicate slopes. Max-Age has
the lowest regret at every horizon, and its fitted slope across the plotted
horizons is close to the reference exponent. This paired comparison shows
that the improvement comes from how the source is chosen rather than a change in
channel learning.

\subsection{Coupled Multi-Source Model}

The coupled experiment uses \(M=K=3\) and eight horizons from \(12000\) to
\(64000\). The service rates follow phase-shifted sinusoidal paths with
\(\alpha=0.3\). Each curve averages 16 service-rate paths and 32 transmission
runs per path. The proposed scheduler uses the empirical cap \(Q=32\), which
is smaller than the conservative sufficient cap in
Algorithm~\ref{alg:capped_optimistic_finite_state}. All tunable parameters are selected at \(T=64000\) using a separate set of simulation runs and are then fixed for all evaluation horizons.

\begin{figure}[!t]
\centerline{\includegraphics[width=\simulationfigurewidth]{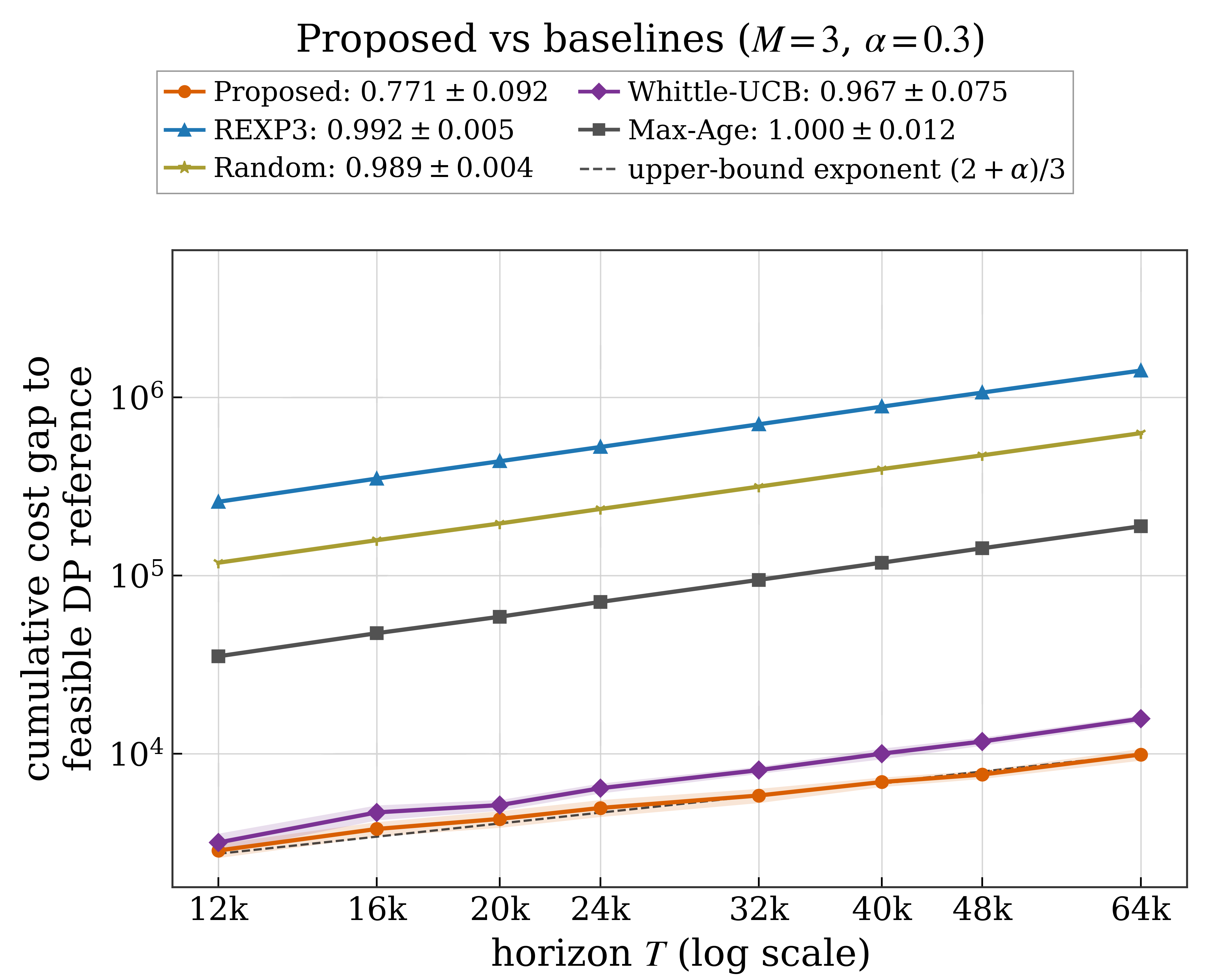}}
\caption{Coupled baseline comparison at \(M=3\) and
\(\alpha=0.3\). The proposed scheduler has the smallest cost gap to
the feasible finite-horizon reference. Shaded bands show the mean \(\pm1.96\)
standard errors across the 16 service-rate trajectories; the legend reports slopes fitted to the
three largest horizons, with \(\pm\) one environment-level bootstrap standard
error.}
\label{fig:coupled-baselines}
\end{figure}

Figure~\ref{fig:coupled-baselines} compares the proposed scheduler with four
baselines. REXP3 learns time-varying channels without using the ages
\cite{besbes2014stochastic}. Max-Age and Random choose sources without
learning and are commonly used in AoI scheduling
\cite{kadota2018scheduling}. Whittle-UCB uses the exact unreliable-channel
index \(A+q\,A(A-1)/2\), where \(A\) is the current age and \(q\) is the
optimistic service-rate estimate~\cite{tripathi2024whittle}.
Because the exact uncapped oracle is infeasible at these horizons, we solve a
finite-horizon capped DP with zero terminal value and run its policy on the
actual uncapped ages. We use cap \(64\) for \(T\le32000\) and \(48\)
otherwise. Each plotted gap is the learner's cost minus the cost of this
feasible rollout, so it is a lower estimate of the true regret. Subtracting
the optimal cost of a capped DP with cap \(96\) from the learner's actual cost
gives an upper estimate of the true regret. Across the eight
horizons, the two estimates differ by at most seven percent of the plotted
gap. The proposed scheduler has the smallest gap
at every horizon and substantially improves over the closest learning-based
state-aware baseline. Its fitted slope is \(0.771\), with an environment-level
bootstrap standard error of \(0.092\), and is numerically consistent with the $(2 + \alpha)/3$ exponent in the proved upper bound. The cost gap of REXP3, which ignores
the ages, grows nearly linearly, while Max-Age and Random have substantially
larger gaps. The results show the benefit of combining channel learning with
age-aware scheduling.

\begin{figure}[!t]
\centerline{\includegraphics[width=\simulationfigurewidth]{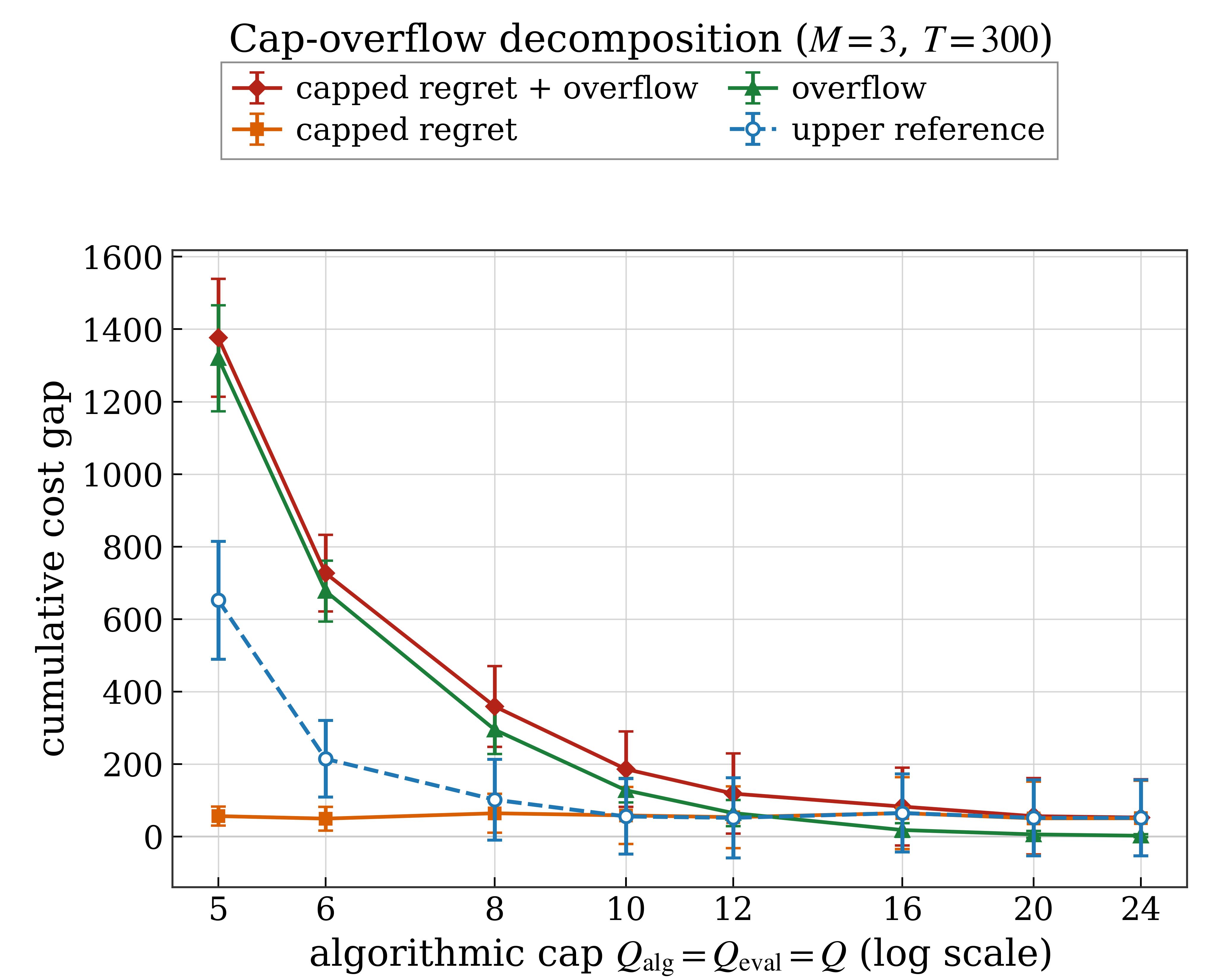}}
\caption{Coupled cap-overflow decomposition. The planning and evaluation
caps are increased together; the resulting reduction comes mainly from the
overflow term.}
\label{fig:coupled-cap-decomposition}
\end{figure}

Figure~\ref{fig:coupled-cap-decomposition} varies the planning and
evaluation cap together. The four curves separate the learning error inside
the capped model, the cost from actual ages above the cap, their sum, and an
upper reference obtained from an optimal capped DP with a larger cap. This is
a separate short-horizon experiment. The overflow
falls quickly with the cap, while the capped learning term remains stable and
the total approaches the upper reference. The figure illustrates the
one-sided capping decomposition: a moderate cap makes overflow negligible
without enlarging the capped learning error.

\section{Conclusion}
\label{sc:conclusion}
We study AoI scheduling when channel service rates are unknown and drift
over time. For a single source, we relate channel regret to AoI regret and
use this relation to obtain lower and upper bounds that agree up to
logarithmic factors. For the decoupled multi-source model, we extend the same learning
principle to multiple ages and show how Max-Age source scheduling can be
combined with REXP3. We also provide BoB versions for these two models when the
variation budget is unknown.

In the coupled multi-source model, choosing a source also chooses its channel, so channel
learning and age-aware scheduling must be solved together. The capped
optimistic scheduler handles this problem as an MDP while retaining the
original finite-horizon oracle as the comparator. Simulations are consistent
with the predicted horizon scaling in the single-source and decoupled multi-source models and show
a clear advantage for state-aware scheduling in the coupled multi-source models. Developing
scalable algorithms and matching lower bounds for the coupled system remains for future work.

\bibliographystyle{IEEEtran}
\bibliography{IEEEabrv,ref}

\renewcommand{\appendixname}{Appendix}
\appendices
\normalsize
\setlength{\abovedisplayskip}{2pt plus 1pt minus 0.5pt}
\setlength{\abovedisplayshortskip}{2pt plus 1pt minus 0.5pt}
\setlength{\belowdisplayskip}{2pt plus 1pt minus 0.5pt}
\setlength{\belowdisplayshortskip}{2pt plus 1pt minus 0.5pt}
\setlength{\jot}{1pt}
\renewcommand{\thesubsection}{\thesection.\arabic{subsection}}
\renewcommand{\thesubsubsection}{\thesubsection.\arabic{subsubsection}}
\renewcommand{\theHequation}{app.\thesection.\arabic{equation}}
\makeatletter
\def\thesubsectiondis{\thesectiondis.\arabic{subsection}.}
\def\thesubsubsectiondis{\thesectiondis.\arabic{subsection}.\arabic{subsubsection}.}
\makeatother

\setcounter{equation}{0}
\renewcommand{\theequation}{\thesection.\arabic{equation}}

\section{Single-Source System Proofs}

The single-source system analysis uses one common age-gap recursion in both
directions. For the lower bound, every oracle-only success is charged
only in the next slot. For the upper bound, the same recursion leaves a
learner-age factor; this age is truncated, and its tail is controlled by the
uniform lower bound~$p$ on every transmission-success probability. Lemmas~\ref{lem:single_lower_transfer} and~\ref{lem:single_upper_transfer} are then combined with the standard drifting-bandit bounds to obtain the AoI regret bounds.

\subsection{Proof of Lemma~\ref{lem:single_lower_transfer}}
\label{app:single_lower_transfer}

\begin{proof}
Let $A^\pi(t)$ and $A^\star(t)$ be the learner and oracle ages at the
beginning of slot~$t$, and let $X_t^\pi,X_t^\star\in\{0,1\}$ be their
success indicators in that slot. Under
Remark~\ref{rem:coupled_channel_states}, the oracle uses a channel whose
service rate is no smaller than the service rate of the learner's selected
channel. Hence, for the same $U_t\sim\operatorname{Unif}[0,1]$,
\[
X_t^\pi=\mathbb{I}\{U_t\le \mu_{k_t^\pi}(t)\}
\le
\mathbb{I}\{U_t\le \max_k\mu_k(t)\}=X_t^\star.
\]

For either trajectory, a success in slot $t$ resets the next-slot age to one,
whereas a failure increases the current age by one. Hence
\[
\begin{aligned}
A^\pi(t+1)&=1+(1-X_t^\pi)A^\pi(t),\\
A^\star(t+1)&=1+(1-X_t^\star)A^\star(t).
\end{aligned}
\]
Define
\[
G_t:=A^\pi(t)-A^\star(t),\qquad
\Delta_t:=X_t^\star-X_t^\pi.
\]
Subtracting the two age recursions and adding and subtracting
$(1-X_t^\star)A^\pi(t)$ gives
\begin{align}
G_{t+1}
&=(1-X_t^\pi)A^\pi(t)-(1-X_t^\star)A^\star(t)\notag\\
&=(1-X_t^\star)\bigl(A^\pi(t)-A^\star(t)\bigr)
 +(X_t^\star-X_t^\pi)A^\pi(t)\notag\\
&=(1-X_t^\star)G_t+A^\pi(t)\Delta_t.
\label{eq:app-single-gap}
\end{align}

We also have $A^\pi(t)\ge A^\star(t)$ for every~$t$. This follows by
induction: the two ages are equal at $t=1$; whenever the learner succeeds,
the oracle succeeds as well, and an oracle-only success can only reduce the
oracle age. Thus $G_t\ge0$. Since $1-X_t^\star\ge0$ and
$A^\pi(t)\ge1$, \eqref{eq:app-single-gap} implies
\[
G_{t+1}\ge \Delta_t.
\]
Therefore, an oracle-only success in slot $t$ creates at least one unit of
AoI gap in slot $t+1$.

Because the construction in Remark~\ref{rem:coupled_channel_states}
preserves the marginal law of each trajectory, the original AoI regret may
be evaluated under this joint construction. Since $G_1=0$,
\begin{align*}
R^\pi(T)
&=\sum_{t=1}^{T}\mathbb E[G_t]
 =\sum_{t=1}^{T-1}\mathbb E[G_{t+1}]\\
&\ge \sum_{t=1}^{T-1}\mathbb E[\Delta_t].
\end{align*}
For a fixed learner choice $k_t^\pi$, Remark~\ref{rem:coupled_channel_states}
gives
\begin{align*}
\mathbb E[\Delta_t\mid k_t^\pi]
&=\Pr\!\left(U_t\le \max_k\mu_k(t)\right)
  -\Pr\!\left(U_t\le \mu_{k_t^\pi}(t)\right)\\
&=\max_k\mu_k(t)-\mu_{k_t^\pi}(t).
\end{align*}
Taking expectation over the learner's random choice therefore yields
\[
\sum_{t=1}^{T}\mathbb E[\Delta_t]=R_{\rm ch}^\pi(T),
\]
and
\[
R^\pi(T)
\ge R_{\rm ch}^\pi(T)-\mathbb E[\Delta_T]
\ge R_{\rm ch}^\pi(T)-1.
\]
The slot-$T$ term is not charged because the slot-$T$ outcome changes the age
only in slot~$T+1$, which lies outside the regret horizon. Since
$R_{\rm ch}^\pi(T)=\Omega(g(T))$ and $g(T)$ grows with $T$, subtracting one
does not change the order.
\end{proof}

\subsection{Proof of Theorem~\ref{th:single-lower}}
\label{app:single_lower_bound}

\begin{proof}
The drifting-bandit lower bound in~\cite[Th.~1]{besbes2014stochastic}
provides an admissible Bernoulli service-rate sequence for which
\[
R_{\rm ch}^\pi(T)=\Omega\!\left((KV_T)^{1/3}T^{2/3}\right).
\]
The hard service-rate sequence in that construction can be shifted and scaled into the
interval $[p,P]$; this changes only numerical constants and preserves the
variation and regret orders.
Lemma~\ref{lem:single_lower_transfer} transfers this channel-regret lower
bound to AoI regret without changing its order.
\end{proof}

\subsection{Proof of Lemma~\ref{lem:single_upper_transfer}}
\label{app:single_upper_transfer}

\begin{proof}
The proof has three steps. First, the age-gap recursion reduces AoI regret to
terms of the form $A^\pi(t)\Delta_t$. Second, the learner age is truncated at
a deterministic threshold. Third, the remaining age tail is bounded by the
probability of consecutive learner failures.

\proofstep{1}{Reduce AoI regret to oracle-only successes}
Fix all random quantities determined before the slot-$t$ transmission outcome
is drawn. At that point $G_t$ is fixed, and the oracle succeeds with
probability at least~$p$. Therefore
\[
\mathbb E[(1-X_t^\star)G_t]\le (1-p)\mathbb E[G_t].
\]
Taking expectation in \eqref{eq:app-single-gap} gives
\[
\mathbb E[G_{t+1}]
\le (1-p)\mathbb E[G_t]
   +\mathbb E[A^\pi(t)\Delta_t].
\]
Summing this inequality for $t=1,\ldots,T$ gives
\[
\sum_{t=1}^{T}\mathbb E[G_{t+1}]
\le (1-p)\sum_{t=1}^{T}\mathbb E[G_t]
 +\sum_{t=1}^{T}\mathbb E[A^\pi(t)\Delta_t].
\]
Since $G_1=0$ and $G_{T+1}\ge0$,
\[
R^\pi(T)=\sum_{t=1}^{T}\mathbb E[G_t]
\le \sum_{t=1}^{T}\mathbb E[G_{t+1}].
\]
Substituting this into the preceding inequality and rearranging yields
\begin{equation}
R^\pi(T)
\le \frac{1}{p}\sum_{t=1}^{T}
\mathbb E[A^\pi(t)\Delta_t].
\label{eq:app-single-upper-core}
\end{equation}
Thus only the learner age multiplying each oracle-only success remains to be
controlled.

\proofstep{2}{Truncate the learner age}
Fix an integer $Q\ge1$. Since $0\le\Delta_t\le1$,
\[
A^\pi(t)\Delta_t
\le Q\Delta_t+(A^\pi(t)-Q)^+.
\]
The first term charges at most $Q$ for each oracle-only success. The second
term is nonzero only when the learner age is larger than~$Q$.

\proofstep{3}{Bound the age tail}
For an integer $r\ge1$, the event $A^\pi(t)\ge r$ requires the immediately
preceding $r-1$ learner transmissions to have failed; any success among those
slots would have reset the age to one. If fewer than $r-1$ preceding slots
exist, the event is impossible. In every slot, regardless of the learner's
channel choice, the conditional failure probability is at most $1-p$.
Applying this bound successively over the preceding slots gives
\[
\Pr\{A^\pi(t)\ge r\}\le(1-p)^{r-1}.
\]
For the nonnegative integer-valued variable $(A^\pi(t)-Q)^+$, the tail-sum
identity gives
\begin{align*}
\mathbb E[(A^\pi(t)-Q)^+]
&=\sum_{s=1}^{\infty}\Pr\{(A^\pi(t)-Q)^+\ge s\}\\
&=\sum_{r=Q+1}^{\infty}\Pr\{A^\pi(t)\ge r\}\\
&\le\sum_{r=Q+1}^{\infty}(1-p)^{r-1}
 =\frac{(1-p)^Q}{p}.
\end{align*}
Substituting the truncation bound into
\eqref{eq:app-single-upper-core} and using
$\sum_t\mathbb E[\Delta_t]=R_{\rm ch}^\pi(T)$ gives
\begin{equation}
R^\pi(T)
\le \frac{Q}{p}R_{\rm ch}^\pi(T)
 +\frac{T(1-p)^Q}{p^2}.
\label{eq:app-single-upper-trunc}
\end{equation}
For $T\ge2$, choose
\[
Q=\left\lceil
\frac{2\log T}{-\log(1-p)}
\right\rceil.
\]
The case $T=1$ is immediate. Because
$Q\ge 2\log T/[-\log(1-p)]$,
\[
(1-p)^Q
=\exp\!\bigl(Q\log(1-p)\bigr)
\le \exp(-2\log T)=T^{-2}.
\]
Hence the second term in \eqref{eq:app-single-upper-trunc} is $O(T^{-1})$,
while $Q=O(\log T)$ and the first term is
$O(\log T\,R_{\rm ch}^\pi(T))$. If
$R_{\rm ch}^\pi(T)=O(g(T))$, the stated result follows.
\end{proof}

\subsection{Proof of Theorem~\ref{th:rexp3}}

\begin{proof}
For the restart length in Algorithm~\ref{alg:rexp3-channel},
\cite[Th.~2]{besbes2014stochastic} gives
\[
R_{\rm ch}^\pi(T)
=O\!\left((K\log K\,V_T)^{1/3}T^{2/3}\right).
\]
Lemma~\ref{lem:single_upper_transfer} adds only a logarithmic factor in~$T$,
so
\[
R^\pi(T)=\widetilde O\!\left((KV_T)^{1/3}T^{2/3}\right).
\]
\end{proof}

\section{Decoupled Multi-Source System Proofs}
\label{app:transfer_decoupled}
\setcounter{equation}{0}

For the lower bound, we first fix the learner's realized success--failure sequence and replace only its source schedule with Max-Age. The oracle-only successes are then inserted in reverse chronological order so that the marginal contribution of each insertion is well defined even when the affected time intervals overlap. For the upper bound, we reverse the comparison and delete the oracle-only successes, bounding for each deletion both the largest sum-AoI gap created and the time until the gap is cleared. Lemmas~\ref{lem:decoupled_lower_transfer} and~\ref{lem:decoupled_upper_transfer} are then combined with the standard drifting-bandit bounds to obtain the AoI regret bounds.

\subsection{Proof of Lemma~\ref{lem:decoupled_lower_transfer}}
\label{app:decoupled_lower_transfer}

\begin{proof}
The proof uses Proposition~\ref{prop:decoupled_separation}, proved in Appendix~\ref{app:decoupled_max_age}. We first represent a
Max-Age state by its most recent successful-update times. We then insert the
oracle-only successes one at a time and compute the marginal AoI reduction of
each insertion.

\proofstep{1}{Represent a Max-Age state by update times}
For a Max-Age trajectory, let $L_m(s)$ be the slot of the most recent
successful update of source~$m$ strictly before the beginning of slot~$s$;
set $L_m(s)=0$ until source~$m$ has its first successful update. Since a
success in slot~$u$ gives age one in slot~$u+1$,
\[
A_m(s)=s-L_m(s).
\]
Therefore
\begin{equation}
\sum_{m=1}^M A_m(s)
=Ms-\sum_{m=1}^M L_m(s).
\label{eq:app-ma-timestamps}
\end{equation}
The source with largest age is the source with smallest $L_m(s)$. Thus a
Max-Age success in slot~$s$ replaces the smallest current update time by~$s$.
Starting from $M$ zeros, an induction over successes shows that the multiset
$\{L_m(s)\}_{m=1}^M$ is exactly the set of the $M$ most recent success slots,
with zeros added when fewer than $M$ successes have occurred.

For a binary success sequence $x=(x_1,\ldots,x_T)$, let $C_{\rm MA}(x)$
denote the cumulative sum-AoI obtained by applying Max-Age to that sequence.
Proposition~\ref{prop:decoupled_separation} gives, pathwise,
\[
C^\pi\ge C_{\rm MA}(X^\pi),
\qquad
C^\star=C_{\rm MA}(X^\star),
\]
where the second equality uses the oracle characterization in
Proposition~\ref{prop:decoupled_separation}. Hence
\[
R^\pi(T)
\ge
\mathbb E\!\left[C_{\rm MA}(X^\pi)-C_{\rm MA}(X^\star)\right].
\]
It is therefore enough to compare two Max-Age trajectories, one driven by the
learner success sequence and one by the oracle success sequence. This
replacement is only an analytical comparison after the complete success
sequence has been fixed; it does not modify the original learner or require
its later channel choices to remain unchanged.

\proofstep{2}{Insert oracle-only successes in reverse time order}
The oracle success sequence contains every learner success and possibly some
additional oracle-only successes. Starting from the learner sequence, insert
these additional successes from the latest one to the earliest one. When an
oracle-only success in slot $t$ is inserted, every later oracle-only success
has already been inserted and remains fixed. The marginal reduction caused
by the current insertion can therefore be charged to this insertion alone.

Consider an insertion with
\[
M\le t\le T-M.
\]
Fix $1\le r\le M$ and examine the state at the beginning of slot~$t+r$.
There are at most $r-1$ success slots strictly between $t$ and $t+r$.
Consequently, after inserting~$t$, the timestamp~$t$ is still among the $M$
most recent timestamps that determine the state in slot~$t+r$.

Before inserting~$t$, at least $M-r+1$ entries among these $M$ timestamps are
no later than $t-1$. Write these entries in nondecreasing order as
$u_1\le\cdots\le u_{M-r+1}\le t-1$. Positive success timestamps are distinct,
and any missing timestamps are zeros. Since at most one success can occur in
each slot,
\[
u_1\le t-(M-r+1).
\]
The insertion of~$t$ displaces the oldest timestamp, which is no larger than
$u_1$. If fewer than $M$ previous successes exist, the displaced entry is a
zero; because $t\ge M$, the same upper bound still holds. Thus the timestamp
sum in \eqref{eq:app-ma-timestamps} increases by at least

\[
t-\bigl[t-(M-r+1)\bigr]=M-r+1.
\]
Since sum-AoI is $Ms$ minus the timestamp sum, the sum-AoI in slot $t+r$
is reduced by at least $M-r+1$.

\proofstep{3}{Sum the marginal reductions}
One interior oracle-only success therefore reduces cumulative sum-AoI by at
least
\[
\sum_{r=1}^M(M-r+1)
=M+(M-1)+\cdots+1
=\frac{M(M+1)}{2}.
\]
The insertion costs telescope: if $C^{(j)}$ is the cost after the first $j$
insertions, then
$C^{(0)}-C^{(J)}=\sum_{j=1}^{J}(C^{(j-1)}-C^{(j)})$.
Hence the marginal reductions add even when their affected slot intervals
overlap. It follows that
\[
R^\pi(T)
\ge \frac{M(M+1)}{2}
\sum_{t=M}^{T-M}\mathbb E[\Delta_t].
\]
The omitted slots consist of the first $M-1$ slots and the final $M$ slots,
including slot $T$. Since $0\le\Delta_t\le1$,
\begin{align*}
\sum_{t=M}^{T-M}\mathbb E[\Delta_t]
&=R_{\rm ch}^\pi(T)
 -\sum_{t=1}^{M-1}\mathbb E[\Delta_t]
 -\sum_{t=T-M+1}^{T}\mathbb E[\Delta_t]\\
&\ge R_{\rm ch}^\pi(T)-(2M-1).
\end{align*}
Therefore
\[
R^\pi(T)
\ge \frac{M(M+1)}{2}
\bigl(R_{\rm ch}^\pi(T)-(2M-1)\bigr).
\]
If $R_{\rm ch}^\pi(T)=\Omega(g(T))$ and $g(T)$ dominates the boundary term
$O(M)$---in particular, under the paper's convention that the model
dimensions are fixed as $T$ grows---the additive term does not change the
order. Hence
\[
R^\pi(T)=\Omega(M^2g(T)).
\]
\end{proof}

\subsection{Proof of Theorem~\ref{th:decoupled_lb}}

\begin{proof}
The channel choices induced by any causal decoupled policy form a causal
$K$-armed bandit policy. This remains true when the channel rule uses the
current ages, because those ages are determined by earlier source choices,
channel choices, transmission outcomes, and policy randomization. The
standard drifting-bandit lower bound in
\cite[Th.~1]{besbes2014stochastic} therefore gives
\[
R_{\rm ch}^\pi(T)
=\Omega\!\left((KV_T)^{1/3}T^{2/3}\right)
\]
for a suitable admissible service-rate sequence.
Lemma~\ref{lem:decoupled_lower_transfer} then yields
\[
R^\pi(T)=\Omega\!\left(M^2(KV_T)^{1/3}T^{2/3}\right).
\]
\end{proof}

\subsection{Proof of Proposition~\ref{prop:decoupled_separation}}
\label{app:decoupled_max_age}

\begin{proof}
We first fix an arbitrary success--failure sequence and compare only source
schedules. We then compare channel choices using
Remark~\ref{rem:coupled_channel_states}.

\proofstep{1}{Compare sorted age vectors for a fixed success sequence}
Let
\[
(A^{\rm MA}(t))^\downarrow=(a_1,\ldots,a_M),\qquad
(A^\rho(t))^\downarrow=(b_1,\ldots,b_M),
\]
where the entries are sorted in nonincreasing order. We prove by induction
that
\[
(A^{\rm MA}(t))^\downarrow
\preceq (A^\rho(t))^\downarrow
\]
for every $t$, where $\preceq$ denotes componentwise order.

The two sorted vectors are equal at $t=1$. Suppose $a_i\le b_i$ for every
$i$ at the beginning of slot~$t$. If slot~$t$ is a failure, every coordinate
in both systems increases by one, so the order is preserved.

Suppose instead that slot $t$ is a success. Max-Age resets its largest
coordinate $a_1$. After the common one-slot increment and sorting,
\[
(A^{\rm MA}(t+1))^\downarrow
=(a_2+1,\ldots,a_M+1,1).
\]
Suppose policy $\rho$ resets the coordinate with rank $r$. Then
\[
(A^\rho(t+1))^\downarrow
=(b_1+1,\ldots,b_{r-1}+1,b_{r+1}+1,\ldots,b_M+1,1).
\]
The first comparison
\[
(a_2+1,\ldots,a_M+1,1)
\preceq (b_2+1,\ldots,b_M+1,1)
\]
follows from $a_i\le b_i$. For the second comparison, deleting an arbitrary
entry $b_r$ from the sorted list $(b_1,\ldots,b_M)$ leaves a vector whose
\(j\)-th entry is either $b_j$ or $b_{j+1}$; both are at least $b_{j+1}$.
Therefore
\[
(b_2+1,\ldots,b_M+1,1)
\preceq (A^\rho(t+1))^\downarrow.
\]
This proves the induction step. Consequently, Max-Age has no larger sum-AoI
in any slot and hence minimizes cumulative sum-AoI pathwise for the fixed
success--failure sequence.

\proofstep{2}{Compare channel service rates}
Under Remark~\ref{rem:coupled_channel_states}, replacing a selected channel
by one with a larger service rate can only turn some failures into successes.
For a sorted Max-Age state, a failure gives
\[
(a_1+1,\ldots,a_M+1),
\]
while a success gives
\[
(a_2+1,\ldots,a_M+1,1).
\]
The success state is componentwise no larger: for the first $M-1$
coordinates, $a_{j+1}\le a_j$, and the last coordinate satisfies
$1\le a_M+1$. An induction over slots therefore shows that adding successes
cannot increase a Max-Age trajectory. Because the source choice does not
affect the success probability in the decoupled model, the oracle combines
Max-Age source scheduling with a channel of largest service rate.
\end{proof}

\subsection{Proof of Lemma~\ref{lem:decoupled_upper_transfer}}
\label{app:decoupled_upper_transfer}

\begin{proof}
The learner belongs to $\Pi_{\rm MA}$, and
Proposition~\ref{prop:decoupled_separation} shows that the oracle also uses
Max-Age. We can therefore compare both systems through the
successful-update-time representation in \eqref{eq:app-ma-timestamps}. For
one oracle-only success, we control two quantities: the largest sum-AoI gap
created by removing that success and the time until later learner successes
remove the gap.

\proofstep{1}{Isolate one oracle-only success}
Start from the oracle success sequence and delete oracle-only successes one at
a time, from the latest one to the earliest one. Consider deleting an
oracle-only success in slot~$t$. All earlier oracle-only successes are still
present, while all later oracle-only successes have already been deleted.
Thus the two marginal trajectories are identical before the slot-$t$ outcome
and use the same success--failure sequence after slot~$t$.

Let $H_t$ be the distance from $t$ to the oracle's $M$-th previous success,
using time~$0$ when fewer than $M$ previous successes exist. In the Max-Age
timestamp list, the success at~$t$ replaces the oldest of the previous $M$
entries. Hence deleting this success decreases the timestamp sum, and
therefore increases sum-AoI, by at most~$H_t$ immediately after slot~$t$.

This marginal gap cannot increase later. Indeed, the timestamp list with the
slot-$t$ success is componentwise no smaller than the list without it. A
common failure leaves both lists unchanged, while a common later success
replaces the smallest timestamp in each list by the same new timestamp. The
timestamp-sum difference is therefore preserved or reduced. Thus the
marginal sum-AoI gap remains at most~$H_t$.

\proofstep{2}{Clearing time}
The marginal contribution of the deleted success at $t$ disappears after the
learner obtains $M$ further successes. At that time, the $M$ update times that
determine both marginal Max-Age states are all later than $t$, so the presence
or absence of the success at $t$ no longer affects either state.

Let $W_1,\ldots,W_M$ be the successive waiting times for these learner
successes. Before each success, every selected channel succeeds with
probability at least~$p$, regardless of earlier outcomes. Consequently,
conditionally on all quantities already determined, $\mathbb E[W_j]\le1/p$.
By linearity of expectation, the conditional expected time to collect the
$M$ successes is at most $M/p$. No independence between $H_t$ and the future
clearing time is needed. Conditioning first and then averaging gives an
expected cumulative contribution of at most
$(M/p)\mathbb E[\Delta_tH_t]$ for slot~$t$.

The successive deletion costs telescope, so summing over all possible
oracle-only successes gives
\begin{equation}
R^\pi(T)
\le \frac{M}{p}\sum_{t=1}^{T-1}
\mathbb E[\Delta_tH_t].
\label{eq:app-dec-upper-core}
\end{equation}

\proofstep{3}{Tail of the gap size}
For $r\ge8M/p$, the event $H_t>r$ means that the oracle has fewer than $M$
successes in the preceding $r$ slots. In each of those slots, the event
$U_s\le p$ in Remark~\ref{rem:coupled_channel_states} guarantees an oracle
success, because every oracle-selected service rate is at least~$p$. The
number of these guaranteed successes is $\operatorname{Bin}(r,p)$. Therefore
\begin{align*}
\Pr\{H_t>r\}
&\le \Pr\{\operatorname{Bin}(r,p)<M\}\\
&\le \Pr\{\operatorname{Bin}(r,p)\le pr/2\}\\
&\le e^{-pr/8}.
\end{align*}
The second line uses $M\le pr/2$, and the last line is the standard Chernoff
lower-tail bound.

Fix an integer $W\ge8M/p$. Since $0\le\Delta_t\le1$,
\[
\Delta_tH_t\le W\Delta_t+(H_t-W)^+.
\]
The tail-sum identity and the preceding exponential bound give
\begin{align*}
\mathbb E[(H_t-W)^+]
&=\sum_{r=W}^{\infty}\Pr\{H_t>r\}\\
&\le\sum_{r=W}^{\infty}e^{-pr/8}
 =\frac{e^{-pW/8}}{1-e^{-p/8}}.
\end{align*}
Hence
\begin{align*}
\sum_{t=1}^{T-1}\mathbb E[\Delta_tH_t]
&\le W\sum_{t=1}^{T-1}\mathbb E[\Delta_t]
 +\sum_{t=1}^{T-1}\mathbb E[(H_t-W)^+]\\
&\le W R_{\rm ch}^\pi(T)
 +\frac{T e^{-pW/8}}{1-e^{-p/8}}.
\end{align*}
Combining this display with \eqref{eq:app-dec-upper-core} gives
\[
R^\pi(T)
\le \frac{MW}{p}R_{\rm ch}^\pi(T)
 +\frac{MT}{p(1-e^{-p/8})}e^{-pW/8}.
\]
Choose
\[
W=\left\lceil\frac{8}{p}(M+3\log T)\right\rceil.
\]
Then $W=O(M+\log T)$ for fixed $p$, and
\[
T e^{-pW/8}
\le T e^{-(M+3\log T)}
=e^{-M}T^{-2}.
\]
Therefore
\[
R^\pi(T)
=O\!\left(M(M+\log T)R_{\rm ch}^\pi(T)\right)+O(MT^{-2}).
\]
If $R_{\rm ch}^\pi(T)=O(g(T))$, the lemma follows.
\end{proof}

\subsection{Proof of Theorem~\ref{th:decoupled_ub}}

\begin{proof}
Max-Age changes only the selected source and does not change the success
probability of the channel selected by REXP3. Thus the REXP3 channel-regret
bound is the same as in the single-source model. Applying
Lemma~\ref{lem:decoupled_upper_transfer} gives
\[
R^\pi(T)
=\widetilde O\!\left(M^2(KV_T)^{1/3}T^{2/3}\right).
\]
\end{proof}

\section{BoB Proofs}
\label{app:bob_proofs}

The BoB proof is first carried out at the channel layer. For a fixed candidate restart window, REXP3 contributes a learning term and a within-window variation term. Restarting the inner learner at the BoB group boundaries incurs only a learning cost of the same order. Outer EXP3 then compares the candidate windows through their group success fractions. The same inner--outer decomposition is used in the original BoB analysis~\cite{cheung2019learning}, its extension to non-stationary RL~\cite{cheung2020reinforcement}, and the CUCB-BoB analysis for non-stationary combinatorial semi-bandits~\cite{chen2021combinatorial}. Combining the inner- and outer-layer regret bounds and applying Lemmas~\ref{lem:single_upper_transfer} and~\ref{lem:decoupled_upper_transfer} gives the corresponding AoI regret bounds.

\subsection{Proof of Corollary~\ref{cor:single-bob}}
\label{app:single_bob_proof}

\begingroup
\setcounter{equation}{0}
\renewcommand{\theequation}{C.\arabic{equation}}
\renewcommand{\theHequation}{bob.C.\arabic{equation}}
\begin{proof}
Let $G_B$ be the group length and define
\[
N_B:=\left\lceil\frac{T}{G_B}\right\rceil,
\qquad
J_B:=|\mathcal W|,
\qquad
N_F:=\left\lfloor\frac{T}{G_B}\right\rfloor.
\]
Thus $N_F$ is the number of full groups; the possible last group has fewer
than $G_B$ slots.

For a full group $g$ and candidate window $w_j$, let $Y_g(j)\in[0,1]$ be the
expected success fraction of a fresh inner run that uses $w_j$ throughout
that group. The service-rate sequence in the group is fixed before the outer
choice is sampled, so $Y_g(j)$ is a valid expected reward for outer arm~$j$.

Let $q_{j,g}>0$ be the probability that outer EXP3 selects $w_j$, let $j_g$
be the selected index, and let $\bar Y_g$ be the observed success fraction in
that group. Use the importance-weighted estimate
\[
\widehat Y_{j,g}
:=\frac{\bar Y_g}{q_{j,g}}\mathbb{I}\{j=j_g\}.
\]
Conditioned on the outer probabilities,
\begin{align*}
\mathbb E[\widehat Y_{j,g}]
&=q_{j,g}\,
  \mathbb E\!\left[\frac{\bar Y_g}{q_{j,g}}\,\middle|\,j_g=j\right]\\
&=Y_g(j).
\end{align*}
Hence the estimate is unbiased. The standard EXP3 comparison therefore gives
\begin{equation}
\max_{j\in[J_B]}\sum_{g=1}^{N_F}Y_g(j)
-\mathbb E\!\left[\sum_{g=1}^{N_F}Y_g(j_g)\right]
\le C\sqrt{N_BJ_B\log J_B}.
\label{eq:app-bob-outer}
\end{equation}
We may use $N_B$ rather than $N_F$ on the right because $N_F\le N_B$.

\proofstep{1}{Cost of a fixed restart window}
Fix $w\in\mathcal W$. If REXP3 uses this restart window over the whole
horizon, its channel regret satisfies
\begin{equation}
B_w
\le C\left[
T\sqrt{\frac{K\log K}{w}}+wV_T
\right].
\label{eq:app-bob-fixed}
\end{equation}
The first term is the repeated EXP3 learning cost: there are at most
$\lceil T/w\rceil$ restart segments, and one segment of length at most $w$
has regret $O(\sqrt{wK\log K})$. The second term is the loss from replacing
the dynamic best channel inside each segment by one fixed channel; a
one-slot change in the service vector affects at most the next $w$ slots,
which gives $O(wV_T)$ after summing all changes.

The BoB wrapper also restarts the inner learner at every group boundary.
Because every candidate satisfies $w\le G_B$, a full-group boundary splits
at most one fixed-window segment and therefore creates at most one additional
segment of length at most~$w$. Across the $N_F$ full groups, these extra
segments contribute at most
\begin{align*}
CN_F\sqrt{wK\log K}
&\le C\frac{T}{G_B}\sqrt{wK\log K}\\
&\le CT\sqrt{\frac{K\log K}{w}},
\end{align*}
where the last inequality is equivalent to $w\le G_B$. Thus the additional
group-boundary restarts are absorbed into the first term of
\eqref{eq:app-bob-fixed}.

\proofstep{2}{Convert the outer EXP3 bound to channel regret}
For a full group $\mathcal I_g$, write
$\mu_t^*:=\max_{k\in[K]}\mu_k(t)$. If candidate~$j$ is used in this group,
its expected channel regret is
\[
\sum_{t\in\mathcal I_g}\mu_t^*-G_BY_g(j).
\]
The first term is independent of the candidate. Consequently, the regret
difference between two candidates is exactly $G_B$ times the difference of
their expected success fractions. Multiplying
\eqref{eq:app-bob-outer} by $G_B$ gives the outer-learning contribution
$CG_B\sqrt{N_BJ_B\log J_B}$.
The final incomplete group has fewer than $G_B$ slots. Since per-slot channel
regret is at most one, charging this group directly costs at most $CG_B$.
Combining the fixed-window and outer costs,
\begin{align}
R_{\rm ch}^{\pi_{\rm BoB}}(T)
&\le C\min_{w\in\mathcal W}
\left[
T\sqrt{\frac{K\log K}{w}}+wV_T
\right]\notag\\
&\quad+CG_B\sqrt{N_BJ_B\log J_B}+CG_B.
\label{eq:app-bob-master}
\end{align}

\proofstep{3}{Compare the dyadic grid with the balancing window}
Without the restriction $w\in\mathcal W$, the two terms in
\eqref{eq:app-bob-fixed} are equal at
\[
T\sqrt{\frac{K\log K}{w^*}}=w^*V_T,
\qquad
w^*:=(K\log K)^{1/3}\left(\frac{T}{V_T}\right)^{2/3}.
\]
The algorithm uses $G_B=\Theta(\sqrt{KT})$, and the dyadic candidate set has
$J_B=O(\log T)$ elements.

If $w^*\le G_B$, the dyadic grid contains a candidate $w$ with
$w^*/2\le w\le w^*$. Replacing $w^*$ by this candidate changes the decreasing
learning term and the increasing variation term by only constant factors.
Therefore
\[
\min_{w\in\mathcal W}B_w
=
\widetilde O\!\left(
(KV_T)^{1/3}T^{2/3}
\right).
\]

Moreover, since $N_B=O(T/G_B)$,
\begin{align*}
G_B\sqrt{N_BJ_B\log J_B}
&=
O\!\left(
\sqrt{TG_BJ_B\log J_B}
\right)\\
&=
\widetilde O\!\left(K^{1/4}T^{3/4}\right),
\end{align*}
The direct $G_B=O(K^{1/2}T^{1/2})$ term is no larger than this outer term in
the nontrivial regime $K\le T$.

If $w^*>G_B$, choose the largest candidate $w=G_B$. The learning term is
decreasing in $w$ and the variation term is increasing in $w$; since they are
equal at $w^*$ and $G_B<w^*$,
\[
G_BV_T
<
T\sqrt{\frac{K\log K}{G_B}}.
\]
Hence
\[
B_{G_B}
\le C T\sqrt{\frac{K\log K}{G_B}}
=\widetilde O\!\left(K^{1/4}T^{3/4}\right),
\]
and the outer term has the same order.

Substituting both cases into \eqref{eq:app-bob-master} yields
\[
R_{\rm ch}^{\pi_{\rm BoB}}(T)
=
\widetilde O\!\left(
(KV_T)^{1/3}T^{2/3}
+
K^{1/4}T^{3/4}
\right).
\]
Lemma~\ref{lem:single_upper_transfer} converts this channel-regret bound into
the claimed single-source AoI-regret bound.
\end{proof}
\endgroup

\subsection{Proof of Corollary~\ref{cor:decoupled-bob}}
\label{app:decoupled_bob_proof}

\begin{proof}
In the decoupled model, Max-Age changes only the selected source. It does not
change the success probability of the channel selected by the inner REXP3
learner. Therefore every channel-layer step in the proof of
Corollary~\ref{cor:single-bob} is unchanged, and
\[
R_{\rm ch}^{\pi_{\rm BoB}}(T)
=
\widetilde O\!\left(
(KV_T)^{1/3}T^{2/3}+K^{1/4}T^{3/4}
\right).
\]
Applying
Lemma~\ref{lem:decoupled_upper_transfer} gives
\[
R^\pi(T)
=\widetilde O\!\left(
M^2\left[(KV_T)^{1/3}T^{2/3}+K^{1/4}T^{3/4}\right]
\right).
\]
\end{proof}

\section{Coupled Multi-Source System MDP Proof}
\label{app:coupled}
Three objects remain distinct throughout the proof: the actual uncapped process, its capped cost representation, and a stationary capped MDP used only as an analytical reference. The benchmark remains the original finite-horizon oracle; it is not replaced by an oracle optimized for capped ages.

The argument has four parts. First, we reduce uncapped regret to capped regret plus learner overflow. Second, we establish global and local structural bounds for the bias on the bias function of a stationary capped AoI MDP. Third, we compare both the learner and the original oracle with the same frozen stationary MDP in each block. Finally, we use the local bias bound to rule out persistent starvation and control the uncapped-age overflow. Substituting the algorithm parameters then proves Theorem~\ref{thm:coupled}. We use the failure state $f(s)$ and reset state $r_i(s)$ from Section~\ref{subsec:coupled-scheduler}, and write
\[
	c(s):=\sum_{m=1}^M s_m.
\]
Throughout this appendix, $C>0$ denotes a generic numerical constant independent of $M,Q,T,V_T$; its value may change from line to line. We use $c_0>0$ for the numerical constant in exponential-decay terms. Dependence on the fixed service-rate bounds $p$ and $P$ is displayed when relevant. \setcounter{equation}{0}

\subsection{Capped comparison and ACOE Setup}

Let $\Pi_{\rm unc}$ be the original oracle-policy class for the coupled multi-source system. A policy in this class knows the complete service-rate path and observes the actual uncapped ages, but chooses its current action before observing the current transmission outcome. The original finite-horizon oracle $\pi^\star$ is an optimizer over this class for the uncapped cumulative AoI. When a policy in $\Pi_{\rm unc}$ is evaluated below with a capped cost, its actions are still generated from the original uncapped information; only the ages used to evaluate the auxiliary cost are capped. For a policy $\rho$, define its state in the capped process componentwise by
\[
	S_Q^\rho(t) :=\bigl(\min\{A_1^\rho(t),Q\},\ldots, \min\{A_M^\rho(t),Q\}\bigr),
\]
and define its capped cumulative cost by
\[
	J_Q^\rho(T) :=\mathbb E_\rho\!\left[\sum_{t=1}^T c(S_Q^\rho(t))\right].
\]
Set
\[
	R_Q^\pi(T) :=J_Q^\pi(T)-\inf_{\rho\in\Pi_{\rm unc}}J_Q^\rho(T),
\]
and define the learner overflow
\[
	\Delta_Q^\pi(T) :=\sum_{t=1}^T\sum_{m=1}^M \mathbb E_\pi[(A_m^\pi(t)-Q)^+].
\]
For every age value,
\[
	A_m^\pi(t) =\min\{A_m^\pi(t),Q\}+(A_m^\pi(t)-Q)^+,
\]
so the learner's uncapped cumulative cost equals $J_Q^\pi(T)+\Delta_Q^\pi(T)$. For every oracle policy $\rho$, capping can only reduce its cost, so its uncapped cost is at least $J_Q^\rho(T)$. Taking the infimum over the same original policy class gives
\begin{equation}
	R^\pi(T)\le R_Q^\pi(T)+\Delta_Q^\pi(T).
	\label{eq:app-cap-comparison}
\end{equation}
The learner side is an exact decomposition, whereas the oracle side uses only the inequality that capping cannot increase cost. This asymmetry is why \eqref{eq:app-cap-comparison} is a one-sided upper bound and why no oracle overflow term is needed. Both costs inside $R_Q^\pi(T)$ are capped, but the comparator policy class remains $\Pi_{\rm unc}$; no new capped oracle is introduced.

We now temporarily set aside the nonstationary learner--oracle comparison and establish structural properties of an arbitrary stationary capped MDP. These properties will later be used in both the capped-regret and overflow analyses. Fix a stationary service vector $\nu=(\nu_1,\ldots,\nu_M)\in[p,P]^M$. Let $h_\nu$ be a bias function of the corresponding capped MDP and let $\lambda^\star(\nu)$ be its optimal average cost. Operationally, a difference $h_\nu(s)-h_\nu(s')$ measures the transient excess capped cost caused by starting from $s$ rather than $s'$, after the common long-run average cost has been removed. Normalize
\[
	h_\nu(\mathbf 1)=0,\qquad \mathbf 1:=(1,\ldots,1).
\]
This is valid because adding a constant to a bias function does not change the Bellman equation. We first verify that the standard average-cost optimality equation (ACOE) applies to the capped MDP. Under any stationary policy, \(Q\) consecutive failures drive every capped state to \((Q,\ldots,Q)\) with positive probability, since each selected channel fails with probability at least \(1-P>0\). A recurrent class is closed, so if it contains a state, every state reachable from that state with positive probability must remain in the same class. Therefore every recurrent class must contain the common state \((Q,\ldots,Q)\). Two disjoint recurrent classes cannot both contain this state, so the induced Markov chain has a single recurrent class and the capped MDP is unichain. By standard finite-state unichain average-cost MDP theory \cite[Secs.~8.3--8.4]{puterman1994markov}, the optimal average cost \(\lambda^\star(\nu)\) is independent of the initial state, and there exists a bias function \(h_\nu\) satisfying the ACOE. For state \(s\) and action \(i\), define the one-step continuation value
\[
	P_\nu h_\nu(s,i) :=\nu_i h_\nu(r_i(s))+(1-\nu_i)h_\nu(f(s)),
\]
and the reset benefit
\[
	D_i(\nu,s) :=h_\nu(f(s))-h_\nu(r_i(s)).
\]
The ACOE is
\begin{align}
	\lambda^\star(\nu)+h_\nu(s) & =c(s)+\min_{i\in[M]}P_\nu h_\nu(s,i)\notag        \\
	                     & =c(s)+h_\nu(f(s)) -\max_{i\in[M]}\nu_iD_i(\nu,s).
	\label{eq:app-acoe}
\end{align}
Since
\[
	P_\nu h_\nu(s,i)=h_\nu(f(s))-\nu_iD_i(\nu,s),
\]
and the first term does not depend on the action, minimizing the continuation value is equivalent to maximizing \(\nu_iD_i(\nu,s)\). Thus an average-cost optimal action at \(s\) maximizes this score; with the tie-breaking in Section~\ref{subsec:coupled-scheduler}, this is exactly the selector used by the capped planner. The same ACOE also immediately gives, for every action \(i\in[M]\),
\begin{equation}
	\lambda^\star(\nu)+h_\nu(s) \le c(s)+P_\nu h_\nu(s,i),
	\label{eq:app-bellman-ineq}
\end{equation}
because the minimum in \eqref{eq:app-acoe} is no larger than the continuation value of any prescribed action \(i\). Equality holds whenever \(i\) is optimal at \(s\). We will repeatedly use these two forms below: \eqref{eq:app-acoe} for the action selected by the capped planner, and \eqref{eq:app-bellman-ineq} for comparison with an arbitrary prescribed action.

\subsection{AoI-Specific Bias Structure}

The proof uses two complementary properties of the capped-MDP bias. Lemma~\ref{lem:coupled-global-bias} bounds the bias difference between arbitrary capped states, which controls the service-rate estimation and within-block drift terms in the capped-regret analysis. Lemma~\ref{lem:coupled-local-bias} gives a sharper bound when only one source age increases by one. This bound is used to compare the reset benefits of old and recently updated sources, rule out persistent starvation, and control actual-age overflow. To state the global bound, define the uniform bias span

$$
H_Q :=\sup_{\nu\in[p,P]^M} \max_{s,s'\in[Q]^M}|h_\nu(s)-h_\nu(s')|.
$$

We bound $H_Q$ using the standard connection between bias span and expected hitting times in average-reward MDPs~\cite{jaksch2010near,bartlett2009regal}. A round-robin policy controls the bias difference from an arbitrary state to a common reference state through its expected hitting time, while a synchronous coupling controls the reverse difference and establishes monotonicity. The AoI reset structure then gives an explicit uniform bound.

\begin{lemma}[Uniform global bias span]
	\label{lem:coupled-global-bias}
	For every \(Q\ge M\) and \(\nu\in[p,P]^M\), \(h_\nu(s)\) is nondecreasing in every coordinate, and
	\[
		H_Q\le Cp^{-M}M^2Q.
	\]
\end{lemma}

\begin{proof}
	We compare every state with the common reference
	\[
		s^\circ:=(M,M-1,\ldots,1).
	\]
	A round-robin policy gives one direction of the comparison. A synchronous coupling gives the reverse direction and also proves monotonicity.

	\proofstep{1}{Reach the reference state}
	Run the round-robin schedule $1,2,\ldots,M,1,2,\ldots$. If all $M$ transmissions in one round succeed, the state after the round is exactly $s^\circ$, regardless of the initial state: source $1$ was reset $M-1$ slots ago and therefore has age $M$, while source $M$ was reset in the last slot and has age~$1$. Let $\tau^\circ$ be the first hitting time of $s^\circ$. Each transmission succeeds with probability at least~$p$, so a complete round succeeds with probability at least $p^M$. Repeated conditioning over successive rounds gives
	\[
		\Pr\{\tau^\circ>nM\}\le(1-p^M)^n.
	\]
	Using the tail-sum formula and grouping slots into rounds,
	\begin{align}
		\mathbb E[\tau^\circ] & \le M\sum_{n=0}^{\infty}\Pr\{\tau^\circ>nM\}\notag \\
		                      & \le M\sum_{n=0}^{\infty}(1-p^M)^n =Mp^{-M}.
		\label{eq:app-hit-reference}
	\end{align}
	Apply the ACOE inequality~\eqref{eq:app-bellman-ineq} to the round-robin action. Because the hitting time is random, first stop at the bounded time $\tau_n:=\tau^\circ\wedge n$. For the round-robin action $i_t$, rearranging \eqref{eq:app-bellman-ineq} gives
	\[
		h_\nu(s(t))-P_\nu h_\nu(s(t),i_t) \le c(s(t))-\lambda^\star(\nu).
	\]
	To sum this inequality up to $\tau_n$, multiply the slot-$t$ relation by $\mathbf 1\{t<\tau_n\}$. This indicator is determined by the history before the slot-$t$ transmission outcome is observed. Hence it can be taken through the conditional expectation defining $P_\nu$, and the tower property gives
	\begin{equation}
		\mathbb E_s\!\left[ \sum_{t<\tau_n}P_\nu h_\nu(s(t),i_t) \right] = \mathbb E_s\!\left[ \sum_{t<\tau_n}h_\nu(s(t+1)) \right].
		\label{eq:app-stopped-continuation}
	\end{equation}
	Thus, the left-hand side can be written in terms of successive differences of \(h_\nu\), which telescope to
	\begin{align*}
		 & \bigl[h_\nu(s(0))-h_\nu(s(1))\bigr] +\bigl[h_\nu(s(1))-h_\nu(s(2))\bigr] +\cdots   \\
		 & \qquad +\bigl[ h_\nu(s(\tau^\circ\wedge n-1)) -h_\nu(s(\tau^\circ\wedge n)) \bigr] \\
		 & = h_\nu(s)-h_\nu(s(\tau^\circ\wedge n)).
	\end{align*}
	Therefore,
	\begin{align*}
		h_\nu(s) -\mathbb E_s[ h_\nu(s(\tau^\circ\wedge n)) ] \le \mathbb E_s \left[ \sum_{t<\tau^\circ\wedge n} \bigl(c(s(t))-\lambda^\star(\nu)\bigr) \right].
	\end{align*}
	Letting $n\to\infty$ is justified by dominated convergence. The finite state space makes $h_\nu$ bounded, and $0\le c(s),\lambda^\star(\nu)\le MQ$, so the absolute running-cost sum stopped at $\tau_n$ is bounded by $2MQ\tau^\circ$, which is integrable by \eqref{eq:app-hit-reference}. Since $s(\tau^\circ)=s^\circ$, we obtain
	\begin{align}
		h_\nu(s)-h_\nu(s^\circ) & \le \mathbb E_s \left[ \sum_{t<\tau^\circ} \bigl(c(s(t))-\lambda^\star(\nu)\bigr) \right] \le MQ\,\mathbb E_s[\tau^\circ] \nonumber \\
		                        & \le M^2p^{-M}Q.
		\label{eq:app-global-one-side}
	\end{align}
	The second inequality uses $c(s(t))\le MQ$ and $\lambda^\star(\nu)\ge0$.

	\proofstep{2}{Synchronous coupling and monotonicity}
	Take arbitrary \(x,y\in[Q]^M\). Let \(\bar s(t)\) start from \(y\), and at each slot let $i_t$ be an action that is optimal for the current state $\bar s(t)$. Let \(\widetilde s(t)\) start from \(x\), apply this same action $i_t$, and use the same transmission outcome as the trajectory \(\bar s(t)\). Thus only the first trajectory is required to act optimally; the second trajectory is a coupled comparison process that copies its actions. For each coordinate, a common successful reset makes the two values equal to one. Even if that coordinate is never successfully reset, both values are repeatedly mapped by
	\[
		a\mapsto\min\{a+1,Q\}
	\]
	and therefore eventually coalesce at~$Q$. Thus every coordinate agrees within at most $Q$ slots, irrespective of the transmission outcomes:
	\[
		\tau_{\rm meet} := \inf\{t\ge0:\widetilde s(t)=\bar s(t)\} \le Q.
	\]
	At $\bar s(t)$ the chosen action is optimal, so \eqref{eq:app-bellman-ineq} is an equality. At $\widetilde s(t)$ the copied action need not be optimal, so only the inequality is available. Subtracting the equality from the inequality gives
	\[
		\begin{aligned}
			 & h_\nu(\widetilde s(t))-h_\nu(\bar s(t)) -\bigl[ P_\nu h_\nu(\widetilde s(t),i_t) -P_\nu h_\nu(\bar s(t),i_t) \bigr] \\
			 & \qquad\le c(\widetilde s(t))-c(\bar s(t)).
		\end{aligned}
	\]
	The left-hand side is a one-step difference of bias gaps between the two coupled trajectories. Since $\tau_{\rm meet}\le Q$, we may sum the inequality over $t<\tau_{\rm meet}$ and use the same stopped tower-property argument as in \eqref{eq:app-stopped-continuation}. The continuation terms satisfy
	\[
		\begin{aligned}
			 & \mathbb E\!\left[
			 \sum_{t<\tau_{\rm meet}}
			 \bigl(
			 P_\nu h_\nu(\widetilde s(t),i_t)
			 -P_\nu h_\nu(\bar s(t),i_t)
			 \bigr)
			 \right] \\
			 &\qquad=
			 \mathbb E\!\left[
			 \sum_{t<\tau_{\rm meet}}
			 \bigl(
			 h_\nu(\widetilde s(t+1))
			 -h_\nu(\bar s(t+1))
			 \bigr)
			 \right].
		\end{aligned}
	\]
	Hence the bias gaps telescope from their initial difference to their terminal difference. By the definition of $\tau_{\rm meet}$, $\widetilde s(\tau_{\rm meet})=\bar s(\tau_{\rm meet})$, so the terminal difference is zero. Therefore
	\begin{equation}
		h_\nu(x)-h_\nu(y) \le \mathbb E \left[ \sum_{t<\tau_{\rm meet}} \bigl(c(\widetilde s(t))-c(\bar s(t))\bigr) \right].
		\label{eq:app-sync-coupling}
	\end{equation}
	If $y\ge x$ componentwise, the common updates preserve this order, so $c(\bar s(t))\ge c(\widetilde s(t))$ for every~$t$. The right-hand side of \eqref{eq:app-sync-coupling} is then nonpositive, which gives $h_\nu(x)\le h_\nu(y)$. This proves coordinatewise monotonicity.

	\proofstep{3}{Reverse comparison with the reference state}
	In \eqref{eq:app-sync-coupling}, set $x=s^\circ$ and $y=s'$. Under the common updates, a positive coordinate difference cannot increase: a common age increment preserves or reduces the difference because of the cap, while a common successful reset eliminates it. Thus, for every $t<\tau_{\rm meet}$, {
			\[
				c(\widetilde s(t))-c(\bar s(t)) \le \sum_{m=1}^M(\widetilde s_m(t)-\bar s_m(t))^+ \le \sum_{m=1}^M(s_m^\circ-s'_m)^+.
			\]
		} Together with $\tau_{\rm meet}\le Q$,
	\begin{align}
		h_\nu(s^\circ)-h_\nu(s') & \le Q\sum_{m=1}^M(s_m^\circ-s_m')^+ \le Q\sum_{m=1}^M(s_m^\circ-1) \nonumber \\
		                         & = \frac{M(M-1)}{2}Q.
		\label{eq:app-global-reverse}
	\end{align}
	The second line uses $s'_m\ge1$. Combining \eqref{eq:app-global-one-side} and \eqref{eq:app-global-reverse}, every state $x$ satisfies
	\[
		-\frac{M(M-1)}{2}Q \le h_\nu(x)-h_\nu(s^\circ) \le M^2p^{-M}Q.
	\]
	The difference between the biases of any two states is therefore at most $Cp^{-M}M^2Q$. Taking the supremum over $\nu\in[p,P]^M$ proves the span bound.
\end{proof}

The global span controls comparisons between arbitrary capped states. For the overflow argument, however, we need a sharper bound when only one source age increases by one. We first use a renewal-reward AoI calculation~\cite{kaul2012realtime,yates2021age} to bound the average cost of a cyclic schedule and hence the expected cost of a finite optimal trajectory. We then couple trajectories starting from neighboring states and use a first-passage bias comparison~\cite{bartlett2009regal} to bound the local bias increment by their expected meeting time. The capped AoI dynamics then yield the required bound on this meeting time.

\begin{lemma}[Uniform local bias increment]
	\label{lem:coupled-local-bias}
	Let \(e_j\) be the \(j\)-th unit vector. For every \(\nu\in[p,P]^M\) and every \(s,s+e_j\in[Q]^M\),
	\[
		0 \le h_\nu(s+e_j)-h_\nu(s) \le Cp^{-M}M^2\sqrt Q.
	\]
\end{lemma}

\begin{proof}
	The lower bound follows immediately from the coordinatewise monotonicity in Lemma~\ref{lem:coupled-global-bias}. For the upper bound, we first control the expected cost of a finite optimal trajectory, then couple trajectories starting from the two neighboring states.

	\proofstep{1}{A finite-prefix cost bound}
	To control a finite prefix of an optimal trajectory, we first need an explicit upper bound on the optimal average cost $\lambda^\star(\nu)$. We obtain such a bound from the following feasible cyclic schedule:
	\[
		1,2,\ldots,M,1,2,\ldots.
	\]
	For source \(i\), let \(G_i\) be the number of its attempts between two consecutive successful updates. Then $G_i$ is geometric on $\{1,2,\ldots\}$ with success parameter $\nu_i$, so
	\[
		\mathbb E[G_i]=\frac{1}{\nu_i},\qquad \mathbb E[G_i^2]=\frac{2-\nu_i}{\nu_i^2}.
	\]
	Because source $i$ is attempted once every $M$ slots, the number of slots between its successful updates is $Z_i=MG_i$. Hence
	\[
		\mathbb E[Z_i] = \frac{M}{\nu_i}, \qquad \mathbb E[Z_i^2] = \frac{M^2(2-\nu_i)}{\nu_i^2}.
	\]
	During a renewal interval of length $Z_i$, the beginning-of-slot ages form $1,2,\ldots,Z_i$, so the cycle accumulates $Z_i(Z_i+1)/2$ units of age. The renewal-reward formula therefore gives the average uncapped age of source~$i$ as \cite{kaul2012realtime,yates2021age}
	\begin{align*}
		\frac{\mathbb E[Z_i^2]+\mathbb E[Z_i]}{2\mathbb E[Z_i]} & =\frac{M(2-\nu_i)}{2\nu_i}+\frac12 \le\frac{CM}{p}.
	\end{align*}
	Summing this bound over the \(M\) sources, the cyclic policy has total average uncapped AoI at most \(CM^2/p\). Capping can only reduce this cost, and \(\lambda^\star(\nu)\) is the optimal average cost of the capped MDP. Therefore
	\begin{equation}
		\lambda^\star(\nu) \le \frac{CM^2}{p}.
		\label{eq:app-average-cost-bound}
	\end{equation}
	Now start an optimal stationary capped trajectory $\bar s(t)$ from $s$. Because its action is optimal in every slot, the ACOE relation \eqref{eq:app-bellman-ineq} holds with equality. Summing for $t=0,\ldots,n-1$ and taking expectations gives
\begin{align}
\mathbb E_s \left[ \sum_{t=0}^{n-1} c(\bar s(t)) \right]
&= n\lambda^\star(\nu)+h_\nu(s)-\mathbb E_s[h_\nu(\bar s(n))] \nonumber \\
&\le \frac{CM^2n}{p}+H_Q
\le Cp^{-M}M^2(Q+n).
\label{eq:app-prefix-cost}
\end{align}
	The final line uses Lemma~\ref{lem:coupled-global-bias} and $p^{-1}\le p^{-M}$ for $M\ge1$.

	\proofstep{2}{Couple neighboring states}
	Start a second trajectory \(\widetilde s(t)\) from \(s+e_j\), copy the actions of \(\bar s(t)\), and use the same transmission outcomes. Let
	\[
		\sigma := \inf\{t\ge0:\widetilde s(t)=\bar s(t)\}.
	\]
	Before \(\sigma\), the two trajectories differ only in coordinate \(j\), by one. This difference disappears after a successful reset of source \(j\), or when the two copies reach the cap. Hence
	\[
		\sigma\le Q.
	\]
	Applying \eqref{eq:app-sync-coupling} with $x=s+e_j$ and $y=s$, the one-step cost difference before meeting is exactly one. Hence
	\begin{equation}
		h_\nu(s+e_j)-h_\nu(s) \le \mathbb E[\sigma].
		\label{eq:app-local-to-meeting}
	\end{equation}
	\proofstep{3}{Meeting-time tail}
	The key implication is that a long meeting time forces the lower trajectory to accumulate a large prefix cost in coordinate $j$. We can therefore bound the probability of a long meeting time using the expected prefix-cost bound from Step~1. Fix \(1\le n<Q\). On the event \(\{\sigma>n\}\), source \(j\) cannot have successfully reset during the first \(n\) slots; otherwise the two trajectories would meet. The lower copy also cannot have reached the cap, because then both copies of coordinate $j$ would equal $Q$ and would again meet. Therefore
	\[
		\bar s_j(0)\ge1,\quad \bar s_j(1)\ge2,\quad\ldots,\quad \bar s_j(n-1)\ge n,
	\]
	and hence
	\[
		\sum_{t=0}^{n-1}c(\bar s(t)) \ge \sum_{t=0}^{n-1}\bar s_j(t) \ge \frac{n(n+1)}2 \qquad\text{on }\{\sigma>n\}.
	\]
	Equivalently,
	\[
		\frac{n(n+1)}2\mathbf 1\{\sigma>n\} \le \mathbf 1\{\sigma>n\} \sum_{t=0}^{n-1}c(\bar s(t)).
	\]
	Taking expectations, dropping the indicator on the right-hand side, and then using \eqref{eq:app-prefix-cost} gives
	\begin{equation}
		\Pr(\sigma>n) \le Cp^{-M}M^2 \frac{Q+n}{n(n+1)}.
		\label{eq:app-meeting-tail}
	\end{equation}
	To sum this tail, split at \(n_0:=\lceil\sqrt Q\rceil\); this balances the first \(n_0\) trivial tail terms against the \(Q/n_0\) contribution from the remaining bound. Since \(\sigma\le Q\), the tail-sum identity and \eqref{eq:app-meeting-tail} give {\small
	\[
		\mathbb E[\sigma] =\sum_{n=0}^{Q-1}\Pr(\sigma>n) \le n_0+Cp^{-M}M^2\sum_{n=n_0}^{Q-1}\frac{Q+n}{n(n+1)}.
	\]
	} For the remaining sum,
	\begin{align*}
		\sum_{n=n_0}^{Q-1}\frac{Q+n}{n(n+1)} & = Q\left(\frac{1}{n_0}-\frac{1}{Q}\right) +\sum_{m=n_0+1}^{Q}\frac{1}{m} \\
		                                     & \le \frac{Q}{n_0}+\log\frac{Q}{n_0}.
	\end{align*}
	Since \(n_0=\lceil\sqrt Q\rceil\),
	\[
		\frac{Q}{n_0}\le\sqrt Q, \qquad \log\frac{Q}{n_0}\le\log Q\le\sqrt Q, \qquad n_0\le2\sqrt Q.
	\]
	Therefore
	\[
		\mathbb E[\sigma] \le Cp^{-M}M^2\sqrt Q.
	\]
	Combining this with \eqref{eq:app-local-to-meeting} proves the lemma.
\end{proof}

\subsection{Blockwise Capped Regret}

We next analyze each restart block using a frozen stationary reference MDP and an optimistic planning MDP for each episode. Confidence bounds control estimation error, count doubling controls replanning, and the variation budget controls within-block drift. The analysis combines UCRL2-style confidence and count-doubling arguments~\cite{jaksch2010near}, adaptive-sampling concentration and average-cost perturbation bounds~\cite{ortner2020variational}, and block restarts for non-stationarity~\cite{cheung2020reinforcement}. The AoI bias span $H_Q$ controls the resulting service-rate estimation and drift terms in the capped-regret bound.

\begin{proposition}[Blockwise capped regret]
	\label{prop:coupled-capped}
	For every \(Q\ge M\), \(1\le L\le T\), and \(\delta\in(0,1)\), Algorithm~\ref{alg:capped_optimistic_finite_state} satisfies
	\begin{align*}
		R_Q^\pi(T) & \le CH_QT\sqrt{\frac{M\log(MT/\delta)}{L}} +CH_QLV_T \\
		           & \quad+CH_QM(1+\log L)\frac{T}{L} +MQT\delta.
	\end{align*}
\end{proposition}
The four terms respectively control service-rate estimation, drift within a block, replanning and block boundaries, and failure of the confidence event. The span $H_Q$ converts a service-rate error into a difference in future capped AoI.

\begin{proof}
	Three service vectors appear in this proof and play different roles: the actual nonstationary vector $\mu(t)$, the block-start vector $\mu^b$ that is held fixed only for analysis, and the learner's episode-wise planning vector $q_k$. The stationary MDP with vector $\mu^b$ serves as a common bridge: we compare the original oracle and the learner separately against this same reference and then subtract the two comparisons. Accordingly, the proof has three stages. First, we define the frozen reference and quantify its sensitivity to service-rate perturbations. Second, we compare the original oracle and the learner separately with that reference: the oracle comparison uses the ACOE inequality, while the learner comparison additionally accounts for estimation error and within-block drift. Finally, we sum the one-step comparisons over episodes and blocks and subtract the two bounds.

	\proofstep{1}{Frozen reference model}
	Consider a block $B_b$ with first slot $t_b$, and define
	\[
		\mu^b:=\mu(t_b), \qquad v_b(t):= \sum_{u=t_b}^{t-1} \|\mu(u+1)-\mu(u)\|_\infty.
	\]
	By the triangle inequality,
	\[
		\|\mu(t)-\mu^b\|_\infty\le v_b(t).
	\]
	Moreover, a variation term at slot $u$ contributes to $v_b(t)$ only for later slots $t$ in the same block. There are at most $L$ such slots, so each one-slot variation is counted at most $L$ times in the double sum and
	\begin{equation}
		\sum_b\sum_{t\in B_b}v_b(t)\le LV_T.
		\label{eq:app-drift-double-sum}
	\end{equation}
	\proofstep{2}{Bellman sensitivity}
	For any bias function $h$ whose span is at most $H_Q$,
	\[
		(P_{\nu'}-P_\nu)h(s,i) = (\nu_i'-\nu_i) \bigl(h(r_i(s))-h(f(s))\bigr).
	\]
	Therefore
	\begin{equation}
		|(P_{\nu'}-P_\nu)h(s,i)| \le H_Q\|\nu'-\nu\|_\infty.
		\label{eq:app-transition-sensitivity}
	\end{equation}
	This is the only place where the transition structure enters the perturbation bound: changing $\nu_i$ changes only the probabilities of the reset and failure next states. The first consequence is that the optimal average cost is Lipschitz in the service vector. Let $i_{\nu'}^\star(s)$ denote an optimal stationary action for the MDP with service vector $\nu'$. By the ACOE inequality for $\nu$ and \eqref{eq:app-transition-sensitivity},
\[
    \lambda^\star(\nu)+h_\nu(s)
    \le
    c(s)+P_{\nu'}h_\nu(s,i_{\nu'}^\star(s))
    +H_Q\|\nu-\nu'\|_\infty.
\]
   Averaging this inequality over time under the optimal policy for $\nu'$ and letting the horizon tend to infinity gives
\[
    \lambda^\star(\nu)-\lambda^\star(\nu')
    \le H_Q\|\nu-\nu'\|_\infty,
\]
  since the bounded bias boundary term vanishes. Swapping $\nu$ and $\nu'$ yields
	\begin{equation}
		|\lambda^\star(\nu)-\lambda^\star(\nu')| \le H_Q\|\nu-\nu'\|_\infty.
		\label{eq:app-average-lipschitz}
	\end{equation}
	We also need an order property: increasing service rates should not increase the optimal AoI cost because it shifts transition probability from failure states toward reset states. Formally, if $\nu'\ge\nu$ componentwise, Lemma~\ref{lem:coupled-global-bias} gives $D_i(\nu,s)\ge0$, and hence
	\[
		P_{\nu'}h_\nu(s,i) \le P_\nu h_\nu(s,i)
	\]
	for every action $i$. Applying the standard ACOE policy-comparison argument~\cite[Sec.~8.4]{puterman1994markov} to an optimal stationary policy for $\nu$ gives
	\begin{equation}
		\nu'\ge\nu \quad\Longrightarrow\quad \lambda^\star(\nu')\le\lambda^\star(\nu).
		\label{eq:app-average-monotone}
	\end{equation}
	\proofstep{3}{Oracle comparison}
	Because the oracle cost will be subtracted when regret is formed, we need a lower bound on its capped cost relative to the frozen reference. Fix any $\rho\in\Pi_{\rm unc}$ and write $h_b:=h_{\mu^b}$. If $I_t^\rho$ is the source selected by $\rho$ in slot $t$, the frozen ACOE inequality \eqref{eq:app-bellman-ineq} gives
	\[
		c(S_Q^\rho(t))-\lambda^\star(\mu^b) \ge h_b(S_Q^\rho(t)) - P_{\mu^b}h_b(S_Q^\rho(t),I_t^\rho).
	\]
	The action $I_t^\rho$ need not be optimal for the frozen stationary MDP, which is why an inequality, rather than equality, is used. Adding and subtracting the continuation under the actual service vector gives
	\begin{align*}
		 & h_b(S_Q^\rho(t)) -P_{\mu^b}h_b(S_Q^\rho(t),I_t^\rho)         \\
		 & \quad= h_b(S_Q^\rho(t)) -P_{\mu(t)}h_b(S_Q^\rho(t),I_t^\rho) \\
		 & \qquad+ (P_{\mu(t)}-P_{\mu^b}) h_b(S_Q^\rho(t),I_t^\rho).
	\end{align*}
	By \eqref{eq:app-transition-sensitivity} and $\|\mu(t)-\mu^b\|_\infty\le v_b(t)$, the last term is at least $-H_Qv_b(t)$. Moreover, although $\rho$ may choose $I_t^\rho$ using the uncapped ages, once that action is fixed the next capped state depends on the past only through the current capped state and the transmission outcome. Hence the conditional capped-state transition is still exactly the reset/failure transition with service vector $\mu(t)$:
	\[
		P_{\mu(t)}h_b(S_Q^\rho(t),I_t^\rho) = \mathbb E[ h_b(S_Q^\rho(t+1)) \mid S_Q^\rho(t),I_t^\rho].
	\]
	Therefore,
	\begin{align}
		c(S_Q^\rho(t))-\lambda^\star(\mu^b) & \ge h_b(S_Q^\rho(t))-H_Qv_b(t)\notag                            \\
		                                    & \quad-\mathbb E[ h_b(S_Q^\rho(t+1)) \mid S_Q^\rho(t),I_t^\rho].
		\label{eq:app-oracle-frozen}
	\end{align}
	\proofstep{4}{Confidence bounds for the learner}
	Consider episode \(k\) inside block \(b\). Let \(N_{k,i}\) be the number of previous observations of source \(i\) in the block. For \(N_{k,i}>0\), define the mean service probability at those sampled slots by
	\[
		\bar\mu_{k,i}
		:=
		\frac1{N_{k,i}}
		\sum_{n=1}^{N_{k,i}}
		\mu_i(\tau_{b,i,n}),
	\]
	where \(\tau_{b,i,n}\) is the slot of the \(n\)-th selection of source \(i\) in this block.

	First control the statistical error. Although the sampling times \(\tau_{b,i,n}\) are chosen adaptively, the decision to select source \(i\) is made from the past, before the current transmission outcome is observed. The centered observations are therefore bounded martingale differences with conditional success means \(\mu_i(\tau_{b,i,n})\). Applying the bounded adaptive-sampling concentration argument of \cite[Proof of Lem.~8]{ortner2020variational} and a union bound over all relevant episode starts and sources gives, with probability at least \(1-\delta\),
	\begin{equation}
		|\widehat\mu_{k,i}-\bar\mu_{k,i}|
		\le
		\beta_{k,i}
		\label{eq:app-confidence}
	\end{equation}
	whenever \(N_{k,i}>0\). Set \(\beta_{k,i}=1\) when \(N_{k,i}=0\).

	Next control the drift between the sampled mean and the current service rate. Every sample entering \(\bar\mu_{k,i}\) lies earlier in the same block, with \(\tau_{b,i,n}\le t_k\le t\) for every slot \(t\) of episode \(k\). Hence the total service-rate change from any sampled slot to \(t\) is at most \(v_b(t)\), and
	\[
		|\bar\mu_{k,i}-\mu_i(t)|
		\le
		v_b(t).
	\]

	Combining the statistical and drift errors controls the planning parameter. On the confidence event, \(\widehat\mu_{k,i}+\beta_{k,i}\ge\bar\mu_{k,i}\in[p,P]\), so clipping gives
	\[
		\bar\mu_{k,i}
		\le
		q_{k,i}
		\le
		\widehat\mu_{k,i}+\beta_{k,i}.
	\]
	In particular,
	\[
		q_{k,i}
		\ge
		\bar\mu_{k,i}
		\ge
		\mu_i(t)-v_b(t),
	\]
	while
	\[
		q_{k,i}-\mu_i(t)
		\le
		|\widehat\mu_{k,i}-\bar\mu_{k,i}|
		+
		|\bar\mu_{k,i}-\mu_i(t)|
		+
		\beta_{k,i}.
	\]
	Therefore
	\begin{equation}
		|q_{k,i}-\mu_i(t)|
		\le
		2\beta_{k,i}+v_b(t).
		\label{eq:app-planning-close}
	\end{equation}
	At the episode start \(t_k\), the same lower bound gives
	\begin{equation}
		q_{k,i}
		\ge
		\mu_i^b-v_b(t_k).
		\label{eq:app-planning-lower}
	\end{equation}
	For \(N_{k,i}=0\), these bounds also hold under the definitions \(q_{k,i}=P\) and \(\beta_{k,i}=1\).

	\proofstep{5}{Compare the learner planning model with the frozen model}
	The confidence construction makes \(q_k\) optimistic for the sampled within-block average, not necessarily for the block-start vector \(\mu^b\). Because the service rates may drift between the block start and episode $k$, we cannot directly assert $q_k\ge\mu^b$, so the monotonicity relation \eqref{eq:app-average-monotone} cannot yet be applied. Introduce the auxiliary vector
	\[
		q_k^+:=q_k\vee\mu^b
	\]
	componentwise. It is used only as a comparison bridge: it dominates \(\mu^b\), while \eqref{eq:app-planning-lower} guarantees that raising $q_k$ to $q_k^+$ costs at most the accumulated drift up to the episode start:
	\[
		q_k^+\ge\mu^b, \qquad \|q_k^+-q_k\|_\infty \le v_b(t_k).
	\]
	Applying the Lipschitz bound and then monotonicity,
	\begin{align}
		\lambda^\star(q_k) & \le \lambda^\star(q_k^+)+H_Qv_b(t_k) \nonumber \\
		                   & \le \lambda^\star(\mu^b)+H_Qv_b(t_k).
		\label{eq:app-plan-vs-frozen}
	\end{align}
	Let
	\[
		h_k(s):=h_{q_k}(s),
	\]
	and let \(i=I_t\) be the learner's selected source. Since \(i\) is optimal for the planning model \(q_k\), the ACOE equality gives
	\[
		c(S_Q^\pi(t)) + P_{q_k}h_k(S_Q^\pi(t),i) - h_k(S_Q^\pi(t)) = \lambda^\star(q_k).
	\]
	Adding and subtracting \(P_{\mu(t)}h_k(S_Q^\pi(t),i)\) gives
	\begin{align*}
     & c(S_Q^\pi(t))
     + P_{\mu(t)}h_k(S_Q^\pi(t),i)
     - h_k(S_Q^\pi(t)) \\
     & \qquad=
     \lambda^\star(q_k)
     + (P_{\mu(t)}-P_{q_k}) h_k(S_Q^\pi(t),i).
    \end{align*}
	Using \eqref{eq:app-transition-sensitivity}, \eqref{eq:app-planning-close}, and \eqref{eq:app-plan-vs-frozen},
	\begin{align}
     & c(S_Q^\pi(t))
     + P_{\mu(t)}h_k(S_Q^\pi(t),i)
     - h_k(S_Q^\pi(t))
     \nonumber\\
     &\qquad\le
     \lambda^\star(\mu^b)
     +H_Qv_b(t_k)
     +H_Q\bigl(2\beta_{k,i}+v_b(t)\bigr)
     \nonumber\\
     &\qquad\le
     \lambda^\star(\mu^b)
     +2H_Q\bigl(\beta_{k,i}+v_b(t)\bigr),
     \label{eq:app-learner-one-step}
     \end{align}
	where the last line uses \(v_b(t_k)\le v_b(t)\).

	\proofstep{6}{Count episode boundaries}
	Although the state space contains \(Q^M\) states, action \(i\) always uses the same unknown service probability \(\mu_i(t)\). Hence all observations are pooled into only \(M\) source counters.

	Inside one block, an episode ends when one source count reaches twice its value at the episode start, with \(1\) used in place of a zero count. By the standard count-doubling argument used in UCRL2~\cite{jaksch2010near}, each source counter can trigger at most \(C(1+\log L)\) episode endings within a block. Thus each block contains at most \(CM(1+\log L)\) episodes. Since there are at most \(CT/L\) blocks,
	\begin{equation}
		N_{\rm ep}
		\le
		CM(1+\log L)\frac{T}{L}.
		\label{eq:app-episode-count}
	\end{equation}

	\proofstep{7}{Sum the confidence radii}
	We next sum the radii over all selections. For a source--episode pair with positive starting count, let \(n_{k,i}\) be the number of selections of source \(i\) during episode \(k\), and let \(N_{k,i}\ge1\) be its count at the episode start. By the doubling rule, \(n_{k,i}\le N_{k,i}\). The standard UCRL2 summation bound~\cite[Lem.~19]{jaksch2010near} then gives, for each source \(i\),
	\[
		\sum_k
		\frac{n_{k,i}}{\sqrt{N_{k,i}}}
		\le
		C\sqrt{N_{b,i}},
	\]
	where \(N_{b,i}\) is the total number of selections of source \(i\) within the block. Intuitively, the episode-start counts grow geometrically, so the inverse-square-root contributions sum to the order of the square root of the final count.

	Summing over the \(M\) sources and applying Cauchy--Schwarz,
	\[
		\sum_{i=1}^M\sqrt{N_{b,i}}
		\le
		\sqrt{M\sum_{i=1}^M N_{b,i}}
		\le
		\sqrt{ML}.
	\]
	Multiplying by the confidence factor and summing over at most \(CT/L\) blocks gives
	\[
		CT\sqrt{\frac{M\log(MT/\delta)}{L}}.
	\]

	For a zero-count source, the crude radius \(1\) can be charged only to its first selection in the block: after that selection its count is positive, and the zero-count doubling threshold ends the current episode. Hence these cases contribute at most \(O(M)\) per block. Therefore,
	\begin{equation}
		\sum_{t=1}^T
		\beta_{k(t),I_t}
		\le
		C\left[
		T\sqrt{\frac{M\log(MT/\delta)}{L}}
		+
		M\frac{T}{L}
		\right].
		\label{eq:app-radius-sum}
	\end{equation}

	\proofstep{8}{Sum the learner comparison}
	Define the common frozen reference
	\[
		\Lambda_Q:=\sum_b|B_b|\lambda^\star(\mu^b).
	\]
	This is only an analytical quantity; neither process is assumed to use the stationary policy attaining $\lambda^\star(\mu^b)$. For each episode whose confidence bounds hold at its start, sum \eqref{eq:app-learner-one-step} over the episode and then take expectations. Since $h_k$ is fixed throughout the episode and
\[
\mathbb E\!\left[
P_{\mu(t)}h_k(S_Q^\pi(t),I_t)
\right]
=
\mathbb E\!\left[
h_k(S_Q^\pi(t+1))
\right],
\]
the intermediate bias terms telescope in expectation. If an episode ends in state $s$ and the next episode uses $h_{k+1}$, the two episode-wise telescopes leave a boundary mismatch of the form $h_{k+1}(s)-h_k(s)$. By the common normalization and Lemma~\ref{lem:coupled-global-bias},
	\[
		0\le h_\nu(s)\le H_Q
	\]
	for every capped state, because $s\ge\mathbf1$ componentwise and $h_\nu(\mathbf1)=0$. Hence $|h_{k+1}(s)-h_k(s)|\le H_Q$, so each change of planning bias contributes at most $H_Q$. Using \eqref{eq:app-drift-double-sum}, \eqref{eq:app-episode-count}, and \eqref{eq:app-radius-sum}, the episodes whose confidence bounds hold at their starts contribute, in expectation, at most
\[
    CH_QT\sqrt{\frac{M\log(MT/\delta)}{L}}
    +CH_QLV_T
    +CH_QM(1+\log L)\frac{T}{L}.
\]
If any episode-start confidence bound fails, we instead use the crude bound $MQT$ for the learner's capped cumulative cost. By the union bound, this contributes at most $MQT\delta$ in expectation. Therefore,
	\begin{align}
		J_Q^\pi(T)-\Lambda_Q & \le CH_QT\sqrt{\frac{M\log(MT/\delta)}{L}} +CH_QLV_T\notag \\
		                     & \quad+CH_QM(1+\log L)\frac{T}{L} +MQT\delta.
		\label{eq:app-learner-frozen-expectation}
	\end{align}

	\proofstep{9}{Sum the oracle comparison and form regret}
	For the oracle, sum \eqref{eq:app-oracle-frozen} over one block. The same $h_b$ is used throughout that block, so the intermediate bias terms telescope. The initial-minus-terminal bias is at least $-H_Q$. Thus
	\[
		\mathbb E\!\left[\sum_{t\in B_b}c(S_Q^\rho(t))\right] -|B_b|\lambda^\star(\mu^b)
		\ge -H_Q-H_Q\sum_{t\in B_b}v_b(t).
	\]
	There are at most $\lceil T/L\rceil\le2T/L$ blocks. Summing over blocks and using \eqref{eq:app-drift-double-sum} gives, for every $\rho\in\Pi_{\rm unc}$,
	\begin{equation}
		J_Q^\rho(T)-\Lambda_Q \ge-CH_Q\left(LV_T+\frac{T}{L}\right).
		\label{eq:app-oracle-frozen-sum}
	\end{equation}
	The two comparisons now share the same reference $\Lambda_Q$, which cancels when regret is formed:
	\begin{align*}
		R_Q^\pi(T)
		&=\bigl(J_Q^\pi(T)-\Lambda_Q\bigr)
		-\inf_{\rho\in\Pi_{\rm unc}}
		\bigl(J_Q^\rho(T)-\Lambda_Q\bigr).
	\end{align*}
	Substituting \eqref{eq:app-learner-frozen-expectation} and \eqref{eq:app-oracle-frozen-sum}, and absorbing the smaller $H_QT/L$ term into $H_QM(1+\log L)T/L$, proves the proposition.
\end{proof}

\subsection{Actual AoI Overflow}

Finally, we transfer the capped analysis to the actual uncapped ages. The key step is to show that, once a source becomes sufficiently old, its ACOE selection score dominates the score of any recently reset source. This prevents persistent starvation and turns repeated opportunities for successful service into a geometric tail for the actual age. Throughout this subsection, the post-failure age of source $i$ in capped state $s$ means $f(s)_i=\min\{s_i+1,Q\}$. Related age-priority structures appear in AoI scheduling~\cite{kadota2018scheduling}; here the priority property is derived from the capped-MDP bias function rather than imposed by the policy.

\begin{lemma}[No-starvation and actual AoI overflow]
	\label{lem:coupled-overflow}
	There is a constant \(C_{p,P}\) such that, for
	\[
		Q \ge Q_0 := C_{p,P}M^6p^{-2M},
	\]
	there exists
	\[
		a_Q \le C_{p,P}M^3p^{-M}\sqrt Q < Q
	\]
	such that the selector never serves a source whose post-failure age is at most \(M+1\) while another source has post-failure age larger than \(a_Q\). Consequently,
	\[
		\Delta_Q^\pi(T) \le Cp^{-M}M^2T \exp\!\left( -c_0\frac{p^MQ}{M} \right).
	\]
\end{lemma}

\begin{proof}
	The proof uses the ACOE selector in \eqref{eq:app-acoe} to compare old and recently reset sources. Steps~1--3 establish a scheduling-score separation; Steps~4--5 convert this separation into no starvation and an exponential tail for the actual uncapped ages.

	Every episode plans with a vector in $[p,P]^M$. We therefore prove the score comparison for an arbitrary fixed $\nu\in[p,P]^M$; because the resulting bound is uniform over this set, it remains valid across episode and block boundaries. Fix a capped state $s$. The Bellman selector maximizes
	\[
		\nu_iD_i(\nu,s).
	\]
	\proofstep{1}{Lower-bound the reset benefit of an old source}
	Set
	\[
		x:=f(s), \qquad y:=r_i(s),
	\]
	and let $i_x$ be an optimal action at~$x$. Since $i_x$ attains the Bellman minimum at~$x$, the ACOE gives equality there. At~$y$, the same action need not be optimal, so only the Bellman inequality is available:
	\[
		\lambda^\star(\nu)+h_\nu(x) =c(x)+P_\nu h_\nu(x,i_x),
	\]
	\[
		\lambda^\star(\nu)+h_\nu(y) \le c(y)+P_\nu h_\nu(y,i_x).
	\]
	The equality/inequality distinction is exactly the one established in \eqref{eq:app-bellman-ineq}. Since \(x\ge y\), applying the same action and the same outcome preserves this order, so Lemma~\ref{lem:coupled-global-bias} gives
	\[
		P_\nu h_\nu(x,i_x) \ge P_\nu h_\nu(y,i_x).
	\]
	Subtracting the second ACOE relation from the first and using this continuation order yields
	\begin{align}
		D_i(\nu,s) & = h_\nu(f(s))-h_\nu(r_i(s)) \nonumber \\
		           & \ge c(f(s))-c(r_i(s)) \nonumber       \\
		           & = f(s)_i-1.
		\label{eq:app-reset-lower}
	\end{align}
	The identity $c(f(s))-c(r_i(s))=f(s)_i-1$ is the immediate one-step AoI saving from resetting source $i$ rather than letting it fail. Monotonicity shows that the continuation effect cannot reverse this saving. Thus the reset benefit grows at least linearly with the source's post-failure age.

	\proofstep{2}{Upper-bound the reset benefit of a young source}
	Suppose $f(s)_j\le M+1$. The states $r_j(s)$ and $f(s)$ differ only in coordinate $j$, by
	\[
		d:=f(s)_j-1\le M.
	\]
	There are $d$ unit-coordinate increments between these two endpoint states. Insert the $d-1$ intermediate states and telescope the bias difference:
	\begin{align*}
		D_j(\nu,s) & =h_\nu(f(s))-h_\nu(r_j(s))                                                            \\
		           & =\sum_{\ell=0}^{d-1} \Bigl[ h_\nu(r_j(s)+(\ell+1)e_j) -h_\nu(r_j(s)+\ell e_j) \Bigr].
	\end{align*}
	Each bracket is at most $Cp^{-M}M^2\sqrt Q$ by Lemma~\ref{lem:coupled-local-bias}. Since $d\le M$,
	\begin{equation}
		D_j(\nu,s) \le Cp^{-M}M^3\sqrt Q.
		\label{eq:app-reset-upper}
	\end{equation}
	\proofstep{3}{Separate the scheduling scores}
	Let $C_1$ be the constant in \eqref{eq:app-reset-upper}, and set
	\[
		a_Q
		:=
		\left\lceil
		2C_1Pp^{-(M+1)}M^3\sqrt Q+1
		\right\rceil.
	\]
	This threshold is chosen only to make the worst-case score of an old source---service rate \(p\)---strictly exceed the best-case score of a young source---service rate \(P\). If
	\[
		f(s)_i>a_Q, \qquad f(s)_j\le M+1,
	\]
	then
\begin{align*}
\nu_iD_i(\nu,s)
&\ge p\bigl(f(s)_i-1\bigr) > p(a_Q-1) \\
&> C_1Pp^{-M}M^3\sqrt Q \ge \nu_jD_j(\nu,s).
\end{align*}
	The first line uses \eqref{eq:app-reset-lower}; the last uses $\nu_j\le P$ and \eqref{eq:app-reset-upper}. Thus even if the old source has the smallest admissible service rate $p$ and the young source has the largest admissible rate $P$, the old source has the larger scheduling score once its post-failure age exceeds $a_Q$. Therefore a source with post-failure age at most $M+1$ cannot be selected while a source above $a_Q$ remains unreset. Absorbing fixed \(p,P\) factors into \(C_{p,P}\),
	\[
		a_Q \le C_{p,P}M^3p^{-M}\sqrt Q.
	\]
	The condition $Q\ge C_{p,P}M^6p^{-2M}$ is chosen so that this $O(M^3p^{-M}\sqrt Q)$ threshold is below a fixed fraction of $Q$. Specifically, enlarging $C_{p,P}$ if necessary gives
	\begin{equation}
		a_Q+2 \le \frac Q2.
		\label{eq:app-threshold-below-cap}
	\end{equation}
	Indeed, after dividing by $\sqrt Q$, the required inequality reduces to a constant multiple of $M^3p^{-M}\le\sqrt Q$, whose square is of order $M^6p^{-2M}\le Q$.

	\proofstep{4}{Derive an actual-age tail}
	We now apply the score separation to the capped state obtained from the actual uncapped trajectory. If source \(i\) has gone \(a_Q+2\) slots without a successful reset, then its post-failure age in the capped process is above \(a_Q\). By \eqref{eq:app-threshold-below-cap}, this threshold lies well below \(Q\), so the cap has not erased the distinction between this old source and a recently reset source.

	Consider a subsequent window of \(M\) slots in which every scheduled transmission succeeds. Until source \(i\) is reset, it remains old. If another source \(j\ne i\) succeeds, its age resets to \(1\); during the fewer than \(M\) slots left in the window, its post-failure age remains at most \(M+1\). Step~3 therefore prevents \(j\) from being selected again while \(i\) remains old. Thus, before \(i\) is reset, successful slots in this window must serve distinct sources. Since there are only \(M\) sources, \(M\) successful slots cannot all avoid \(i\). Hence an all-success \(M\)-slot window must reset \(i\). The same argument clears every source that was already old at the beginning of the window.

	The score comparison is uniform over all planning vectors in \([p,P]^M\), so the conclusion remains valid even if the algorithm replans inside the window. Moreover, in every slot the selected source succeeds with probability at least \(p\). Conditional on the entire past, any specified \(M\)-slot window is therefore all-successful with probability at least \(p^M\).

	After source \(i\) first crosses the old-age threshold, partition the subsequent slots into consecutive complete \(M\)-slot windows, which we call clearing opportunities. If \(A_i^\pi(t)>r\), then source $i$ has spent at least $r$ consecutive slots without a successful reset before slot $t$. The first $a_Q+2$ slots are enough to make it old; the remaining $(r-a_Q-2)^+$ slots therefore contain at least
	\[
		\left\lfloor
		\frac{(r-a_Q-2)^+}{M}
		\right\rfloor
	\]
	disjoint complete clearing opportunities. All of them must have failed to be all-successful, because any all-success opportunity would have reset source \(i\). Each opportunity fails with conditional probability at most \(1-p^M\). Repeated conditioning therefore gives
	\begin{equation}
		\Pr(A_i^\pi(t)>r)
		\le
		(1-p^M)^{
		\left\lfloor
		\frac{(r-a_Q-2)^+}{M}
		\right\rfloor
		}.
		\label{eq:app-actual-age-tail}
	\end{equation}
	Independence between clearing opportunities is not needed: after conditioning on all preceding history, the next opportunity still succeeds with probability at least \(p^M\).

	\proofstep{5}{Sum the overflow tail}
	For integer-valued age,
	\[
		(A_i^\pi(t)-Q)^+ = \sum_{r=Q}^{\infty} \mathbf 1\{A_i^\pi(t)>r\}.
	\]
	Taking expectations and summing over sources and slots,
	\begin{equation}
		\Delta_Q^\pi(T) = \sum_{t=1}^T \sum_{i=1}^M \sum_{r=Q}^{\infty} \Pr(A_i^\pi(t)>r).
		\label{eq:app-overflow-tail-sum}
	\end{equation}
	Let
	\[
		\kappa_Q
		:=
		\left\lfloor
		\frac{Q-a_Q-2}{M}
		\right\rfloor.
	\]
	By \eqref{eq:app-threshold-below-cap},
	\[
		\kappa_Q \ge \frac{Q}{2M}-1.
	\]
	For \(r\ge Q\), the exponent in \eqref{eq:app-actual-age-tail} takes each integer value \(k\ge \kappa_Q\) for at most \(M\) consecutive values of \(r\), because the floor increases by one every \(M\) values of \(r\). Hence
	{\small
\begin{align*}
\sum_{r=Q}^{\infty} \Pr(A_i^\pi(t)>r)
&\le M\sum_{k=\kappa_Q}^{\infty}(1-p^M)^k
 = Mp^{-M}(1-p^M)^{\kappa_Q} \\
&\le Mp^{-M}e^{-p^M\kappa_Q}
 \le Cp^{-M}M\exp\!\left(-c_0\frac{p^M Q}{M}\right).
\end{align*}
}
	where the last two inequalities use \(1-x\le e^{-x}\) and \(\kappa_Q\ge Q/(2M)-1\). Substituting this bound into \eqref{eq:app-overflow-tail-sum} and summing over the \(M\) sources and \(T\) slots gives
	\[
		\Delta_Q^\pi(T) \le Cp^{-M}M^2T \exp\!\left( -c_0\frac{p^MQ}{M} \right).
	\]
\end{proof}

\subsection{Proof of Theorem~\ref{thm:coupled}}

\begin{proof}
	We substitute the choices in Algorithm~\ref{alg:capped_optimistic_finite_state} into Proposition~\ref{prop:coupled-capped} and Lemma~\ref{lem:coupled-overflow}.

	\proofstep{1}{Block length}
	By Assumption~\ref{ass:drifting},
	\[
		V_T=\Theta(T^\alpha), \qquad 0<\alpha<1.
	\]
	For fixed \(M\), \(V_T=\Theta(T^\alpha)\) with \(\alpha>0\) implies
	\[
		M^{1/3}\left(\frac{T}{V_T}\right)^{2/3}
		=
		\Theta\!\left(M^{1/3}T^{2(1-\alpha)/3}\right)
		=
		o(T).
	\]
	Hence, for sufficiently large \(T\), the truncation in the definition of \(L\) is inactive and
	\[
		L = \Theta\!\left( M^{1/3} \left(\frac{T}{V_T}\right)^{2/3} \right), \qquad \delta=T^{-3}.
	\]
	Substituting this choice into the estimation and drift terms gives
	\begin{align}
		T\sqrt{\frac{M\log(MT/\delta)}{L}} & = O\!\left( M^{1/3}V_T^{1/3}T^{2/3} \sqrt{\log(MT)} \right),
		\label{eq:app-final-estimation}
		\\
		LV_T                               & = O\!\left( M^{1/3}V_T^{1/3}T^{2/3} \right).
		\label{eq:app-final-drift}
	\end{align}
	Thus the chosen block length balances estimation and drift up to the logarithmic factor. For fixed $M$, the episode-boundary term is asymptotically smaller in the $T,V_T$ scaling:
	\begin{align}
		M(1+\log L)\frac{T}{L} & = O\!\left( M^{2/3}(1+\log L)V_T^{2/3}T^{1/3} \right) \nonumber \\
		                       & = o\!\left( (1+\log(MT))V_T^{1/3}T^{2/3} \right),
		\label{eq:app-final-boundary}
	\end{align}
	because \(V_T=\Theta(T^\alpha)\) with \(\alpha<1\) implies
	\[
		T^{(\alpha-1)/3}\log T\to0.
	\]
	\proofstep{2}{Age cap and confidence failure}
	The chosen cap satisfies the threshold in Lemma~\ref{lem:coupled-overflow}. For fixed \(M,p,P\), \(Q=O(\log T)\), with its logarithmic coefficient chosen so that
	\[
		c_0 \frac{p^MQ}{M}\ge 2\log T.
	\]
	Hence Lemma~\ref{lem:coupled-overflow} gives
	\begin{align}
		\Delta_Q^\pi(T) & \le Cp^{-M}M^2T e^{-2\log T} \nonumber \\
		                & = Cp^{-M}M^2T^{-1}.
		\label{eq:app-final-overflow}
	\end{align}
	Thus the actual-AoI overflow is lower order. Also, since \(\delta=T^{-3}\) and \(Q=O(\log T)\),
	\[
		MQT\delta = O\!\left(\frac{\log T}{T^2}\right),
	\]
	so the confidence-failure contribution is lower order as well.

	\proofstep{3}{Final substitution}
	Lemma~\ref{lem:coupled-global-bias} gives
	\[
		H_Q\le Cp^{-M}M^2Q.
	\]
	Substituting the preceding bounds into Proposition~\ref{prop:coupled-capped} yields
	\[
		R_Q^\pi(T) = O\!\left( p^{-M}M^{7/3}Q (1+\log(MT)) V_T^{1/3}T^{2/3} \right).
	\]
	By \eqref{eq:app-cap-comparison} and \eqref{eq:app-final-overflow},
	\[
		R^\pi(T) \le R_Q^\pi(T)+\Delta_Q^\pi(T)
	\]
	\[
		= O\!\left( p^{-M}M^{7/3}Q (1+\log(MT)) V_T^{1/3}T^{2/3} \right).
	\]
	For fixed \(M,p,P\), and \(Q=O(\log T)\),
	\[
		R^\pi(T) = \widetilde O\!\left( V_T^{1/3}T^{2/3} \right).
	\]
\end{proof}

\end{document}